\documentclass[11pt]{article}
\usepackage{ifthen}

\ifthenelse{\isundefined{\GenerateShortVersion}}
{
\def\LongVersion{}
\def\LongVersionEnd{}
\long\def\ShortVersion#1\ShortVersionEnd{}
}
{
\def\ShortVersion{}
\def\ShortVersionEnd{}
\long\def\LongVersion#1\LongVersionEnd{}
}

\newcommand{\Comment}[1]{\ignorespaces}

\usepackage{amsmath, amssymb}

\usepackage{amsthm}

\usepackage{ifpdf}

\ifpdf
\usepackage[pdftex]{graphicx}
\usepackage[pdftex]{hyperref}
\else
\usepackage[dvips]{graphicx}
\fi

\ifthenelse{\isundefined{\NoGenerateLetterSize}}
{
\ifpdf
\setlength{\pdfpagewidth}{8.5in}
\setlength{\pdfpageheight}{11in}
\else

\fi
}
{}

\usepackage{multirow}

\usepackage[noend]{algorithmic}
\usepackage{algorithm}

\newtheorem{theorem}{Theorem}[section]
\newtheorem{lemma}[theorem]{Lemma}

\newtheorem{corollary}[theorem]{Corollary}
\newtheorem{proposition}[theorem]{Proposition}

\newtheorem*{observation*}{Observation}

\theoremstyle{definition}

\theoremstyle{plain}

\renewcommand{\paragraph}[1]{\vspace{0.05in}\par\noindent\textbf{#1}}

\newcommand{\Alg}[0]{\ensuremath{\mathtt{Alg}}}
\newcommand{\Opt}[0]{\ensuremath{\mathtt{Opt}}}
\newcommand{\leftPoint}[0]{\mathrm{left}}
\newcommand{\rightPoint}[0]{\mathrm{right}}
\newcommand{\Reals}[0]{\mathbb{R}}
\newcommand{\Integres}[0]{\mathbb{Z}}
\newcommand{\Naturals}[0]{\mathbb{N}}

\newcommand{\Oracle}[0]{\mathcal{C}}
\newcommand{\Probability}[0]{\mathbb{P}}
\newcommand{\Expectation}[0]{\mathbb{E}}

\newcommand{\Angles}[1]{\langle{#1}\rangle}

\DeclareMathOperator*{\argmin}{arg\,min}
\DeclareMathOperator*{\argmax}{arg\,max}

\newcommand{\lefty}{\leftPoint}
\newcommand{\righty}{\rightPoint}
\newcommand{\bzone}{\text{\textit{bz}}}
\newcommand{\fzone}{\text{\textit{fz}}}

\newcommand{\Section}[0]{Sec.}

\begin{document}

\title{Space-Constrained Interval Selection\thanks{A preliminary version of this paper appeared in 
the proceedings of ICALP 2010 \cite{EHR12}.}}

\author{
Yuval Emek\thanks{
Technion - Israel Institute of Technology.
Email: \texttt{yemek@ie.technion.ac.il}.}
\and
Magn\'{u}s M.\ Halld\'{o}rsson\thanks{
ICE-TCS, School of Computer Science, Reykjavik University, Iceland.
Email: \texttt{mmh@ru.is}.
Research partially supported by grant 90032021 from the Icelandic Research Fund.}
\and
Adi Ros\'{e}n\thanks{
CNRS and Universit\'e Paris Diderot, France.
Email: \texttt{adiro@liafa.univ-paris-diderot.fr}.
Research partially supported by ANR projects QRAC,  ALADDIN and RDAM.}
}
\date{}

\maketitle

\begin{abstract}
We study streaming algorithms for the interval selection problem:
finding a maximum cardinality subset of disjoint intervals on the line.
A deterministic $2$-approximation streaming algorithm for this problem is
developed, together with an algorithm for the special case of proper intervals,
achieving improved approximation ratio of $3/2$. 
We complement these upper bounds by proving that they are essentially best
possible in the streaming setting:
it is shown that an approximation ratio of $2 - \epsilon$ (or $3 / 2 -
\epsilon$ for proper intervals) cannot be achieved unless the space is linear
in the input size.
In passing, we also answer an open question of Adler and Azar (J.\ Scheduling
2003) regarding the space complexity of constant-competitive randomized
preemptive online algorithms for the same problem.
\end{abstract}

%%%%%%%%%%%%%%%%%%%%%%%%%%%%%%%%%%%%%%%%%%%%%%%%%%%%%%%%%%%%%%%%%%%%%%%%%%%%%%
\section{Introduction}
\label{sec:intro}
%%%%%%%%%%%%%%%%%%%%%%%%%%%%%%%%%%%%%%%%%%%%%%%%%%%%%%%%%%%%%%%%%%%%%%%%%%%%%%
In this paper we consider the \emph{interval selection} problem, namely,
finding a maximum cardinality subset of disjoint intervals from a given
collection of intervals on the real line.
It is well known that this problem has a simple optimal algorithm in the
classical setting when the complete set of intervals is given to the algorithm
\cite{Gavril72}.
Here we study this problem in the \emph{streaming} model
\cite{HRR98,Muthukrishnan05}, where the input is given to the algorithm as
a stream of items (intervals in our case), one at a time, and the algorithm
has a limited memory that precludes storing the whole input.
Yet, the algorithm is still required to output a feasible solution, with a
good approximation ratio.

The motivation for the streaming model stems from applications of managing
very large data sets, such as biological data (DNA sequencing), network
traffic data, and more.
Although some function of the whole data set is to be computed, it is
impossible to store the whole input.
Depending on the setting, different variants of the streaming model have been
considered in the literature, such as the classical streaming model
\cite{HRR98} or the so-called \emph{semi-streaming} model
\cite{Feigenbaum2005}.
Common to all of them is the fact that the space used by the streaming
algorithm is linear in some natural upper bound on the size of the output it
returns (sometimes, a multiplicative polylogarithmic overhead is allowed).

In many problems considered in the streaming literature, the size of the
output is fully determined by some parameter of the
input, and thus, one would typically express the space complexity as a
function of this parameter (cf. \cite{AMS99,FKMSZ08}).
However, in other problems, the size of the output cannot be a priori
expressed that way as it depends on the given instance;
in such settings it is natural to seek a streaming algorithm whose space
complexity is not much larger than the output size of the given instance
(cf. \cite{HalldorssonHLS10}).
Clearly, as long as the computational model of the streaming algorithm is
based on a Turing machine with no distinction between the working tape and the
output tape, the size of the output is an inherent lower bound on the required
space.

In this paper, we consider a setting where the algorithm is given a stream of
real-line intervals, each one defined by its two endpoints, and the
goal is to compute a maximum cardinality subset of disjoint intervals (or an
approximation thereof).
This problem finds many applications, e.g., in resource allocation problems,
and it has been extensively studied in the online and offline settings in many
variants.
We seek algorithms with a good upper bound on the space they use for a given
instance, expressed in terms of the size of the output for that specific
instance.
Typically, we seek algorithms that use space which is at most linear in the
size of the output and yet guarantee a good approximation ratio.

%%%%%%%%%%%%%%%%%%%%%%%%%%%%%%%%%%%%%%%
\paragraph{Related Work.}
%%%%%%%%%%%%%%%%%%%%%%%%%%%%%%%%%%%%%%%
The offline interval selection problem corresponds to finding a maximum
independent set in an interval graph.
An optimal greedy algorithm was discovered early \cite{Gavril72}
and has since been a staple of algorithms textbooks \cite{clrs,Kleinberg05}.
It should be noted that the input can be given in (at least) two different
ways: as an intersection graph with the nodes corresponding to the intervals,
or as a set of intervals given by their endpoints.
This distinction makes little difference in the traditional offline setting,
where switching between these representations can be done efficiently,
but, it can be important in access- or resource-constrained settings.
We choose to study the interval selection problem assuming the latter
representation --- that is, the input is given as a set of intervals --- since
we believe that it makes more sense in applications related to the online and
streaming settings (most previous works on online interval selection make the
same choice).

The study of space-constrained algorithms goes back at least to the 1980 work
of Munro and Paterson on selection and sorting \cite{MunroPaterson80}.
More recently, the streaming model was developed to capture the processing of
massive data-sets that arise in practice \cite{Muthukrishnan05}.
Most streaming algorithms deal with the approximate computation of various
statistics, or ``heavy hitters'', as exemplified by the celebrated paper of
Alon, Matias, and Szegedy \cite{AMS99}.

A number of classic graph theoretic problems have been treated in the
streaming setting, for example, matching problems
\cite{McGregor05,EpsteinLMS10}, diameter and shortest paths
\cite{Feigenbaum2005,FKMSZ08}, min-cut \cite{AG09}, and graph spanners
\cite{FKMSZ08}.
These were mostly studied under the \emph{semi-streaming} model, introduced by
Feigenbaum et al.~\cite{Feigenbaum2005};
in this model, the algorithm is allowed to use $n \log^{O (1)} (n)$ space on
an $n$-vertex graph (i.e., $\log^{O (1)} (n)$ bits per vertex).
Closest to our problem, the independent set problem in general
sparse graphs (and hypergraphs) was studied in the streaming setting by
Halld\'{o}rsson et al.~\cite{HalldorssonHLS10}.
Geometric streaming algorithms have also appeared in recent years,
especially dealing with extent and ranges, such as \cite{Agarwal2010}.

There is a plethora of literature on interval selection in the online
setting.
Some papers capture the problem as a call admission problem on a linear
network, with the objective of maximizing the number (or weight) of accepted
calls.
Awerbuch et al.~\cite{ABFR94} present a strongly $\lceil \log N
\rceil$-competitive algorithm for the problem, where $N$ is the number of
nodes on the line (corresponding to the number of possible interval
endpoints).
This yields an $O(\log \Delta)$-competitive algorithm for the weighted case,
where $\Delta$ is the ratio between the longest to the shortest interval.
On the negative side, they establish a lower bound of
$\Omega(\log N)$ on the competitive ratio of randomized non-preemptive online
interval selection algorithms.
In the context of the real line, this immediately implies that such algorithms
cannot have a competitive ratio that is independent of the length of the
input.
In fact, Bachmann et al.~\cite{BHS10} recently showed that the competitive
ratio of randomized non-preemptive online algorithms for interval selection on
the real line must be linear in the number of intervals in the input.
Preemptive online scheduling has a lower bound of $\Omega(\log \Delta /
\log\log \Delta)$ in the weighted case \cite{CanettiI98}.
In comparison, much better results are possible for preemptive online
algorithms in the unweighted setting:
Adler and Azar~\cite{AdlerAzar03} devise a $16$-competitive algorithm.
One way of easing the task of the algorithm is to assume arrival by time,
i.e., the intervals arrive in order of left endpoints.
This has been treated for different weighted problems
\cite{Woeginger94,Lipton94onlineinterval,EpsteinL10,FPZ08}.

Subsequent to the initial publication of the present results \cite{EHR12},
Cabello and P\'erez-Lanterno \cite{CP15} gave streaming algorithms that estimate
the size of the maximum independent set out of a set of intervals. Their algorithms
give for a general instance a $(2 + \epsilon)$ approximation, and for
unit intervals a $(3/2 + \epsilon)$ approximation, using space
polynomial in $\frac{1}{\epsilon}$ and in $\log n$. They also gave new, simpler
than ours, algorithms for finding the approximated independent set, which in some
cases match our bounds as to the approximation ratio and the space used.

%%%%%%%%%%%%%%%%%%%%%%%%%%%%%%%%%%%%%%%
\paragraph{Our results.}
%%%%%%%%%%%%%%%%%%%%%%%%%%%%%%%%%%%%%%%
We give tight results for the interval selection problem in the streaming
setting.
Our main positive result is a deterministic $2$-approximation streaming
algorithm that uses space linear in the size of the output
(\Section~\ref{section:MainAlgorithm}).
This is complemented by a matching lower bound
(\Section~\ref{section:LowerBounds}), stating that an approximation ratio of $2
- \epsilon$ cannot be obtained by any randomized streaming algorithm with
space significantly smaller than the size of the input (which can be much
larger than the size of the output).
The special case of proper interval collections (i.e., collections of
intervals with no proper containments) is also considered, for which a
deterministic $3/2$-approximation streaming algorithm that uses space linear
in the output size is presented
(\Section~\ref{section:ProperIntervals}) and a matching lower bound on the
approximation ratio is established (\Section~\ref{section:LowerBounds}) for
streams of unit intervals (a special case of proper intervals).
The upper bounds are extended to \emph{multiple-pass} streaming algorithms:
we show that an approximation ratio $1 + 1 / (2 p - 1)$ can be obtained
in $p$ passes over the input (\Section~\ref{section:MultiPass}).

In passing, we also answer an open question posed by Adler and Azar
\cite{AdlerAzar03} in the context of randomized preemptive online algorithms
for the interval selection problem.
Adler and Azar point out that the decisions made by their online algorithm
depend on the whole history (i.e., the input seen so far) and that natural
attempts to remove this dependency seem to fail.
Consequently, they write (using the term ``active call'' for an interval in the
solution maintained by the online algorithm) that
\textit{
``it seems very interesting to find out whether there exist
constant-competitive algorithms where each decision depends only on the
currently active calls and maybe on additional bounded information''}.
We answer this question in the affirmative by slightly modifying our main
algorithm to achieve a randomized preemptive online algorithm that admits
constant competitive ratio and uses space linear in the size of the optimal
solution, rather than the size of the input, as the algorithm of Adler and
Azar does (\Section~\ref{section:Online}).\footnote{
The technique employed in \Section~\ref{section:Online} is based on a
``classify and randomly select'' argument that guarantees that the solution
produced by the online algorithm is a constant approximation of the optimal
solution with constant probability.
Using the technique of \cite{LMPR01} (reformulated as Theorem~4.1 in
\cite{AdlerAzar03}), this can be strengthened to guarantee a constant
approximation with high probability.
}

%%%%%%%%%%%%%%%%%%%%%%%%%%%%%%%%%%%%%%%%%%%%%%%%%%%%%%%%%%%%%%%%%%%%%%%%%%%%%%
\section{Preliminaries}
\label{sec:prelim}
%%%%%%%%%%%%%%%%%%%%%%%%%%%%%%%%%%%%%%%%%%%%%%%%%%%%%%%%%%%%%%%%%%%%%%%%%%%%%%
We think of the real line $\Reals$ as stretching from left to right so that an
\emph{interval} $I$ contains all points between its left \emph{endpoint}
$\leftPoint(I)$ and its right endpoint $\rightPoint(I)$, where $\leftPoint(I)
< \rightPoint(I)$.
Each endpoint can be either \emph{open} (exclusive) or \emph{closed}
(inclusive).
A \emph{half-open} interval has a closed left endpoint and an open right
endpoint.
(This is, perhaps, the natural interval type to use in most resource allocation
applications.)
Observe that the assumption that $\leftPoint(I) < \rightPoint(I)$ implies that
every interval contains an open set (in the topological sense) and that
half-open intervals are always well defined.

The interval related notions of \emph{intersection}, \emph{disjointness}, and
\emph{containment} follow the standard view of an interval as a set of points.
Two intervals $I, J$ \emph{properly} intersect if they intersect without
containment;
$I$ properly contains $J$ if $I$ contains $J$ and $J$ does not contain $I$.
An interval collection $\mathcal{I}$ is said to be \emph{proper} (and the
intervals in the collection, \emph{proper} intervals) if no two intervals in
$\mathcal{I}$ exhibit proper containment.
The \emph{load} of $\mathcal{I}$ is defined to be $\max_{p \in \Reals} |\{ I
\in \mathcal{I} \mid p \in I \}|$.

The \emph{interval selection} problem asks for a maximum cardinality subset of
pairwise disjoint intervals out of a given set $S$ of intervals.
In the streaming model, the input interval set $S$ is considered to be an
ordered set (a.k.a.\ a \emph{stream}) and the intervals arrive one by one
according to that order.
The intervals are specified by their endpoints, where each endpoint is
represented by a bit string of length $b$ (the same $b$ for all endpoints).
This may potentially provide a streaming algorithm with the edge of knowing in
advance some bounds on the number of intervals that will arrive and on
the number of intervals that can be placed between two existing intervals.
However, our algorithms do not take advantage of this extra information and
our lower bounds show that it is essentially useless.
An optimal solution to a given instance $S$ of the interval selection problem
is denoted by $\Opt(S)$.

We may sometimes talk about \emph{segments}, rather than intervals, when we
want to emphasize that the entities under consideration are not necessarily
part of the input.
Given a set $\mathcal{I}$ of intervals, a \emph{component} (or \emph{connected
component}) of $\mathcal{I}$ is a maximal continuous segment in $\bigcup_{I
\in \mathcal{I}} I$.

%%%%%%%%%%%%%%%%%%%%%%%%%%%%%%%%%%%%%%%%%%%%%%%%%%%%%%%%%%%%%%%%%%%%%%%%%%%%%%
\section{The Main Algorithm}
\label{section:MainAlgorithm}
%%%%%%%%%%%%%%%%%%%%%%%%%%%%%%%%%%%%%%%%%%%%%%%%%%%%%%%%%%%%%%%%%%%%%%%%%%%%%%

%%%%%%%%%%%%%%%%%%%%%%%%%%%%%%%%%%%%%%%
\subsection{Overview}
%%%%%%%%%%%%%%%%%%%%%%%%%%%%%%%%%%%%%%%
Given a stream $S$ of intervals, our algorithm maintains a (proper interval)
collection $A \subseteq S$, referred to as the \emph{actual} intervals, from
which the output $\Alg(S) = \Opt(A)$ is taken.
It also maintains a collection $V$ of \emph{virtual} intervals,
where each virtual interval is the intersection of two actual intervals that
existed in $A$ at some point.
The role of the virtual intervals is to prevent undesired intervals from
joining $A$:
an arriving interval $I \in S$ joins $A$ if and only if it does not contain
any currently maintained virtual or actual interval.

Our algorithm is designed to guarantee that each interval $I \in S$ leaves
a \emph{trace} in either $A$ or $V$, namely, there exists some $J \in A
\cup V$ such that $J \subseteq I$.
Moreover, if $I, I' \in A$ properly intersect, then $I \cap I' \in V$.
This essentially means that an arriving interval is rejected if and only if it
contains some previous interval of $S$ or the intersection of two properly
intersecting previous intervals in $S$ that have belonged to $A$.

Following that, it is not too difficult to show that the load of the interval
collection $A$ is at most $2$.
Based on a careful analysis of the structure of the (connected) components in
$A$ and the locations of the virtual intervals within these components and
between them, we can argue that $|V| \leq |A|$.
This immediately yields the desired upper bound on the space of our algorithm
as $|A| \leq 2 \cdot |\Opt(A)|$.
The bound on the approximation ratio essentially stems from the observation
that $|\Opt(S)| \leq |\Opt(A \cup V)|$ (a direct corollary of the fact that
each interval in $S$ leaves a trace in $A \cup V$) and from the invariant that
each actual interval contains at most $2$ virtual intervals.

It is interesting to point out that our algorithm is in fact a
deterministic preemptive online algorithm that maintains a load-$2$ interval
collection (the collection $A$).
Since the main result of Adler and Azar~\cite{AdlerAzar03} also relies on such
an algorithm, one may wonder if the two algorithms can be compared.
Actually, the algorithm of Adler and Azar bases its rejection (and preemption)
decisions on similar conditions:
an arriving interval is rejected if and only if it contains some previous
interval of $S$ or the intersection of two properly intersecting intervals in
$A$.
(Adler and Azar use a different terminology, but the essence is very similar.)
The difference lies in the latter condition:
Whereas the algorithm of Adler and Azar considers only the properly
intersecting intervals that are currently in $A$, our algorithm also
(implicitly) considers properly intersecting intervals that belonged to $A$
in the past and were preempted since.
This seemingly small difference turns out to be crucial as it allows our
algorithm to use much less memory, thus giving rise to an interesting
phenomena:
by remembering extra information (i.e., intersecting intervals that belonged
to $A$ in the past and are not in $A$ anymore), we actually end up using less
memory.

%%%%%%%%%%%%%%%%%%%%%%%%%%%%%%%%%%%%%%%
\subsection{The algorithm}
%%%%%%%%%%%%%%%%%%%%%%%%%%%%%%%%%%%%%%%
Consider a stream $S = (I_1, \dots, I_n)$ of intervals on the real line.
It will be convenient to assume that all endpoints are distinct, i.e., $\{
\leftPoint(I), \rightPoint(I) \} \cap \{ \leftPoint(J), \rightPoint(J) \} =
\emptyset$ for every two intervals $I, J \in S$.
Unless stated otherwise, we will also assume that the intervals mentioned in
this section are closed on both endpoints.
These two assumptions are lifted in
Appendix~\ref{section:LiftingAssumptions}.

Our algorithm, denoted \Alg{}, maintains a collection $A \subseteq S$ of
\emph{actual} intervals and a collection $V$ of \emph{virtual} intervals,
where each virtual interval is realized by endpoints of intervals in $S$.
That is, the virtual interval $I \in V$ satisfies $\{ \leftPoint(I),
\rightPoint(I) \} \subseteq \{ \leftPoint(J), \rightPoint(J) \mid J \in S
\}$.
The algorithm initially sets $A, V \leftarrow \emptyset$.
Then, upon arrival of a new interval $I \in S$, \Alg{} proceeds
according to the policy\footnote{
Note that \Alg{} can be thought of as an online algorithm with preemption with
respect to the set $A$.
} presented in Algorithm~\ref{algorithm:main}.

\begin{algorithm}
\caption{ \label{algorithm:main}
The policy of \Alg{} upon arrival of an interval $I \in S$.
}
\begin{algorithmic}[1]
\IF {$\exists J \in A \cup V$ s.t. $J \subseteq I$}
  \STATE reject $I$ and halt \label{algline:Reject}
\ENDIF
\STATE $A \leftarrow A \cup \{I\}$ \label{algline:AddNewActual}
\FORALL {$J \in A - \{ I \}$ s.t. $J \supseteq
I$} \label{algline:RemoveContainingBegin}
  \STATE $A \leftarrow A - \{J\}$ \label{algline:RemoveContainingA}
\ENDFOR
\FORALL {$J \in V$ s.t. $J \supseteq I$}
  \STATE $V \leftarrow V - \{J\}$ \label{algline:RemoveContainingV}
\ENDFOR \label{algline:RemoveContainingEnd}
\FOR {$p \in \{\leftPoint(I), \rightPoint(I)
\}$} \label{algline:UpdateVBegin}
  \IF {$\exists J \in V$ s.t. $p \in J$}
    \STATE $V \leftarrow V - \{J\} \cup \{ I \cap J
\}$ \label{algline:ReplaceVirtual}
  \ELSIF {$\exists J \in A$ s.t. $p \in J$} \label{algline:Else}
    \STATE $V \leftarrow V \cup \{ I \cap J \}$ \label{algline:AddNewVirtual}
  \ENDIF
\ENDFOR \label{algline:UpdateVEnd}
\FORALL {$J \in A$ and $K \in V$} \label{algline:CleanABegin}
  \IF {$\leftPoint(J) < \leftPoint(K) < \rightPoint(K) < \rightPoint(J)$}
    \STATE $A \leftarrow A - \{J\}$ \label{algline:CleanA}
  \ENDIF
\ENDFOR \label{algline:CleanAEnd}
\end{algorithmic}
\end{algorithm}

The algorithm first verifies that the new interval $I$ does not contain any
currently stored (actual or virtual) interval;
if it does, then the new interval is ignored (rejected).
Therefore, if \Alg{} reaches line~\ref{algline:AddNewActual}, then we
can assume that $I \nsupseteq J$ for any interval $J \in A \cup V$.
Next, in
lines~\ref{algline:RemoveContainingBegin}--\ref{algline:RemoveContainingEnd}
\Alg{} removes all the actual and virtual intervals that contain $I$.
Lines~\ref{algline:UpdateVBegin}--\ref{algline:UpdateVEnd} form the heart
of the algorithm: updating the virtual intervals that remain in $V$.
The idea here is that a virtual interval that intersects with $I$ is
``trimmed'' until it is contained in $I$;
if an actual interval intersects with $I$, then the intersection is introduced
as a new virtual interval.
Finally, any actual interval $J$ that exclusively contains some virtual interval
$K$ (that is, $J$ contains $K$ even if we remove $J$'s endpoints) is removed
from the actual interval collection $A$ in
lines~\ref{algline:CleanABegin}--\ref{algline:CleanAEnd}.

After the last interval $I_n$ is processed, \Alg{} outputs $\Alg(S) =
\Opt(A)$, that is, an optimal subset of the interval collection $A$ (computed,
say, by the greedy left-to-right algorithm).
In the remainder of this section we prove that:
(a) at all times, $|V| \leq |A| \leq 2 \cdot |\Alg(S)|$; and
(b) $|\Alg(S)| \geq |\Opt(S)| / 2$.
Together, we obtain the desired approximation, using space at most constant
times larger than the size of the optimal output.

%%%%%%%%%%%%%%%%%%%%%%%%%%%%%%%%%%%%%%%
\subsection{Analysis}
%%%%%%%%%%%%%%%%%%%%%%%%%%%%%%%%%%%%%%%
Throughout the analysis, we let $1 \leq t \leq n$ denote the time at which
\Alg{} completed processing interval $I_t \in S$;
time $t = 0$ denotes the beginning of the execution.
We refer to the period between time $t - 1$ and time $t$ as \emph{round $t$}.
The stream prefix $(I_1, \dots, I_t)$ is denoted by $S_t$.
The collections $A$ and $V$ at time $t$ are denoted by $A_t$ and $V_t$,
respectively, although, when $t$ is clear from the context, we may omit the
subscript.
We begin by showing that each virtual interval is indeed realized by (at most)
two actual intervals and that the new interval $I$ is not removed immediately
after joining $A$.

\begin{proposition} \label{proposition:VirtualEndpoints}
At any time $t$, we have $\{ \leftPoint(\rho), \rightPoint(\rho) \mid \rho \in
V_t \} \subseteq \{ \leftPoint(\sigma), \rightPoint(\sigma) \mid \sigma \in
S_t \}$.
\end{proposition}
\begin{proof}
By induction on $t$.
The case $t = 0$ is trivial as $V_{0} = \emptyset$.
For time $t > 0$, we observe that any new virtual interval $\rho$ added to
$V$ in round $t$ is either the intersection of two actual
intervals (line~\ref{algline:AddNewVirtual}) or the intersection of an actual
interval and a virtual interval in $V_{t - 1}$
(line~\ref{algline:ReplaceVirtual}).
In the former case, the assertion follows immediately;
in the latter case, the assertion follows by the inductive hypothesis.
\end{proof}

\begin{proposition} \label{proposition:NewActualSurvives}
For every $1 \leq t \leq n$, if \Alg{} reaches line~\ref{algline:AddNewActual}
when processing $I_{t} = I$, then $I \in A_{t}$.
\end{proposition}
\begin{proof}
In line~\ref{algline:AddNewActual}, $I$ is added to $A$ and subsequently, it
can only be removed from $A$ if a virtual interval $\rho$ that is contained in
$I$ but does not have an endpoint in common with $V$ is found
(line~\ref{algline:CleanA}).
Such an interval $\rho$ cannot be in $V_{t - 1}$ since otherwise, $I$ would
have been rejected in line~\ref{algline:Reject}.
The assertion follows since every virtual interval added to $V$ in round $t$
has a common endpoint with $I$.
\end{proof}

Lemma~\ref{lemma:Trace} lies at the core of our analysis:
it states that each interval in $S$ leaves some trace in either $A$ or $V$.
This will be employed later on to argue that $\Alg(S)$ is not much
smaller than $\Opt(S)$.

\begin{lemma} \label{lemma:Trace}
For every interval $I_{t} \in S$ and for every time $t' \geq t$, there exists
some interval $\rho \in A_{t'} \cup V_{t'}$ such that $\rho \subseteq I_{t}$.
\end{lemma}
\begin{proof}
A new coming interval $I$ is added to $A$ in line~\ref{algline:AddNewActual}
unless some interval $\rho \subseteq I$ is found in $A \cup V$.
An actual interval $\rho \in A$ is removed from $A$ only if another actual
interval $I \subseteq \rho$ has just joined $A$
(line:\ref{algline:RemoveContainingA}) or if a virtual interval $\sigma
\subset \rho$ is found in $V$ (line:\ref{algline:CleanA}).
A virtual interval $\rho \in V$ is removed from $V$ only if an actual interval
$I \subseteq \rho$ has just joined $A$ (line:\ref{algline:RemoveContainingV})
or if it is replaced in $V$ by another virtual interval $\sigma \subseteq
\rho$ (line~\ref{algline:ReplaceVirtual}).
The assertion follows.
\end{proof}

%%%%%%%%%%%%%%%%%%%%%%%%%%%%%%%%%%%%%%%
\subsubsection{The structural lemma}
%%%%%%%%%%%%%%%%%%%%%%%%%%%%%%%%%%%%%%%
We now turn to establish our main lemma regarding the updating phase in
lines \ref{algline:UpdateVBegin}--\ref{algline:UpdateVEnd} and the resulting
structure of the interval collections $A$ and $V$.
Lemma~\ref{lemma:Structural} states seven invariants maintained by our
algorithm;
these invariants are then proved simultaneously by induction on $t$,
essentially by straightforward analysis of the policy presented in
Algorithm~\ref{algorithm:main}.

\begin{lemma} \label{lemma:Structural}
For any round $1 \leq t \leq n$, the updating phase satisfies the following
two properties: \\
(P1) If $\rho$ is added to $V$ in round $t$, then $\rho \in V_{t}$. \\
(P2) If $\rho$ and $\sigma$ are added to $V$ in round $t$, then $\rho \cap
\sigma = \emptyset$. \\
Moreover, for any time $0 \leq t \leq n$, the interval collections $A$ and $V$
satisfy the following five properties: \\
(P3) For every $\rho \in A$ and $\sigma \in V$, if $\rho \cap \sigma \neq
\emptyset$, then $\sigma \subset \rho$ with a common endpoint. \\
(P4) For every $\rho, \sigma \in A$, if $\rho \cap \sigma \neq \emptyset$,
then $\rho \cap \sigma \in V$. \\
(P5) Every point $p \in \Reals$ is contained in at most $1$ virtual
interval. \\
(P6) Every point $p \in \Reals$ is contained in at most $2$ actual
intervals. \\
(P7) There do not exist two actual intervals $\rho, \sigma \in A$ such that
$\rho \subseteq \sigma$.
\end{lemma}
\begin{proof}
We first establish (P1) regardless of the other six properties.

\noindent
\textbf{Establishing (P1).}
It is sufficient to show that if $\rho$ is added to $V$ in line
\ref{algline:ReplaceVirtual} or  line \ref{algline:AddNewVirtual} of the execution
for $p = \leftPoint(I)$, then it is not removed from $V$ in
line~\ref{algline:ReplaceVirtual} of the execution for $p = \rightPoint(I)$.
Indeed, if $\rho$ is added to $V$ in the execution for $p = \leftPoint(I)$,
then $\rho = I \cap \sigma$ for some interval $\sigma \in A_{t - 1} \cup V_{t
- 1}$ such that $\leftPoint(I) \in \sigma$.
Since $\sigma$ cannot contain $I$ (as otherwise, it would have been removed in
line \ref{algline:RemoveContainingA} or  line \ref{algline:RemoveContainingV}), it
follows that $\leftPoint(\sigma) < \leftPoint(I) < \rightPoint(\sigma) <
\rightPoint(I)$, so $\rho = [\leftPoint(I), \rightPoint(\sigma)]$.
Therefore, $\rightPoint(I) \notin \rho$ and $\rho$ is not removed from $V$ in
line~\ref{algline:ReplaceVirtual} of the execution for $p = \rightPoint(I)$.

Next, we establish (P2), (P3), (P4), and (P5) simultaneously by induction on
$t$.
The case $t = 0$ is trivial: (P2) holds vacuously, while (P3), (P4), and
(P5) hold as $A_{0} = V_{0} = \emptyset$.
Assume that the four properties hold for $t - 1$ and consider the execution of
\Alg{} upon arrival of interval $I = I_{t}$ for some $1 \leq t \leq n$.

\noindent
\textbf{Establishing (P2).}
As each iteration of the for loop in lines
\ref{algline:UpdateVBegin}--\ref{algline:UpdateVEnd} adds at most one
virtual interval to $V$, we may assume that $\rho$ is added in the
execution for $p = \leftPoint(I)$ and $\sigma$ is added in the execution
for $p = \rightPoint(I)$.
This means that $\rho = I \cap \tau_{\ell}$ and $\sigma = I \cap \tau_{r}$ for
some intervals $\tau_{\ell}, \tau_{r} \in A_{t - 1} \cup V_{t - 1}$ such that
$\leftPoint(I) \in \tau_{\ell}$ and $\rightPoint(I) \in \tau_{r}$.
We argue that $\tau_{\ell}$ and $\tau_{r}$ do not intersect, which implies
that $\rho$ and $\sigma$ do not intersect.

To that end, assume by way of contradiction that they do, and let $\tau_{\cap}
= \tau_\ell \cap \tau_r$.
If both $\tau_{\ell}$ and $\tau_{r}$ are virtual intervals, then we
immediately reach a contradiction due the inductive hypothesis on (P5).
If both $\tau_{\ell}$ and $\tau_{r}$ are actual intervals, which means that
$\rho$ and $\sigma$ are added to $V$ in line~\ref{algline:AddNewVirtual},
then by the inductive hypothesis on (P4), $\tau_{\cap} \in V_{t - 1}$.
By definition, $\tau_{\cap}$ must intersect with $I$.
On the other hand, neither $\leftPoint(I)$ nor $\rightPoint(I)$ can belong to
$\tau_{\cap}$ as otherwise, the else condition in line~\ref{algline:Else}
would not have passed, thus $\tau_{\cap} \subset I$.
But this means that \Alg{} should not have reached
line~\ref{algline:AddNewActual} and in particular, $\rho$ and $\sigma$ would
not have been added to $V$.

So, assume that $\tau_{\ell}$ is actual and $\tau_{r}$ is virtual (the proof
of the converse possibility is identical).
By the inductive hypothesis on (P3), we know that $\tau_{r} \subset \tau_{\ell}$.
But this implies that both endpoints of $I$ belong to $\tau_{\ell}$, namely,
$I \subseteq \tau_{\ell}$, and $\tau_{\ell}$ should have been removed from $A$ in
line~\ref{algline:RemoveContainingA}.

\noindent
\textbf{Establishing (P3).}
Consider some $\rho \in A_{t}$ and $\sigma \in V_{t}$ such that $\rho \cap
\sigma \neq \emptyset$.
If $\rho \in A_{t - 1}$ and $\sigma \in V_{t - 1}$, then the property holds by
the inductive hypothesis.
Assume first that $\rho$ is added to $A$ in round $t$, so $\rho$ is the last
arriving interval $I$.
Notice that $\sigma$ cannot be in $V_{t - 1}$ as this implies that either
(i) $\sigma \subseteq I$, in which case $I$ would have been rejected in
line~\ref{algline:Reject};
(ii) $\sigma \supseteq I$, in which case $\sigma$ would have been removed from
$V$ in line~\ref{algline:RemoveContainingV}; or
(iii) $\sigma$ and $I$ properly intersect, in which case $\sigma$ is removed
from $V$ in line~\ref{algline:ReplaceVirtual}.
Thus, $\sigma$ is added to $V$ in round $t$ either in
line~\ref{algline:ReplaceVirtual} or in line~\ref{algline:AddNewVirtual}.
In both cases, $\sigma$ is contained in $I$ with a common endpoint.

It remains to consider the case in which $\rho \in A_{t - 1}$ and $\sigma$ is
added to $V$ in round $t$.
If $\sigma$ is added to $V$ in line~\ref{algline:ReplaceVirtual}, then it
replaces in $V$ some interval $\tau \in V_{t - 1}$ such that $\sigma \subseteq
\tau$.
Hence, $\tau$ must also intersect with $\rho$ and by the inductive hypothesis,
$\tau \subset \rho$, so $\sigma$ must be contained in $\rho$.
Since $\rho$ is not removed in line~\ref{algline:CleanA}, $\rho$ and $\sigma$
must have a common endpoint.
If $\sigma$ is added to $V$ in line~\ref{algline:AddNewVirtual}, then
$\sigma = I \cap \tau$ for some interval $\tau \in A_{t - 1}$ such that
an endpoint $p$ of $I$ is contained in $\tau$.
The property is established by arguing that $\tau$ and $\rho$ must be the same
interval.

To that end, suppose toward a contradiction that $\tau \neq \rho$.
Assume without loss of generality that $p = \leftPoint(I)$, so
$\leftPoint(\tau) < \leftPoint(I) < \rightPoint(\tau) < \rightPoint(I)$.
Since $\sigma = I \cap \tau = [\leftPoint(I), \rightPoint(\tau)]$ intersects
with $\rho$, both $I$ and $\tau$ must also intersect with $\rho$.
By the inductive hypothesis on (P4), we know that $\sigma_{\cap} = \rho \cap
\tau \in V_{t - 1}$.
We also know that $\sigma_{\cap}$ intersects with $I$ as both $\rho$ and
$\tau$ intersect with $I$.
Since $\tau \nsupseteq I$, it follows that $\sigma_{\cap} \nsupseteq I$, hence
$\sigma_{\cap}$ must still be in $V$ when \Alg{} reaches
line~\ref{algline:UpdateVBegin}.
If $\leftPoint(I) \in \sigma_{\cap}$, then the else condition in
line~\ref{algline:Else} would not have passed and $\sigma$ would not have been
added to $V$ in line~\ref{algline:AddNewVirtual}, so $\leftPoint(I) \notin
\sigma_{\cap}$.
But $\rightPoint(I) \notin \sigma_{\cap}$ as $\rightPoint(I) \notin \tau$, hence
$\sigma_{\cap} \subseteq I$ and $I$ should have been rejected in
line~\ref{algline:Reject}.
In any case, we conclude that $\rho$ and $\tau$ are indeed the same interval.

\noindent
\textbf{Establishing (P4).}
Consider two intersecting intervals $\rho, \sigma \in A_{t}$.
If both $\rho$ and $\sigma$ are also in $A_{t - 1}$, then by the inductive
hypothesis, $\tau = \rho \cap \sigma \in V_{t - 1}$.
If $\tau \notin V_{t}$, then it must have been removed from $V$ either in
line~\ref{algline:RemoveContainingV} because $I \subseteq \tau$, in which case
$I$ is also contained in both $\rho$ and $\sigma$ and they would have been
removed from $A$ in line~\ref{algline:RemoveContainingA}, or in
line~\ref{algline:ReplaceVirtual}, where it is replaced in $V$ by some other
virtual interval $\tau' \subset \tau$ (the strict containment follows from the
distinct endpoints assumption), in which case at least one of the intervals
$\rho$ and $\sigma$ should have been removed in line~\ref{algline:CleanA}.
Therefore, $\tau \in V_{t}$ and the property holds in that case.

So, suppose that $\rho \in A_{t - 1}$, while $\sigma = I$ is added to
$A$ in round $t$.
Since $\rho, I \in A_{t}$, both $\rho$ and $I$ are in $A$ when \Alg{} reaches
line~\ref{algline:UpdateVBegin}, thus they cannot contain each other.
Assume without loss of generality that $\leftPoint(\rho) < \leftPoint(I)
< \rightPoint(\rho) < \rightPoint(I)$.
If $\leftPoint(I)$ does not belong to any virtual interval in $V_{t - 1}$,
then in line~\ref{algline:AddNewVirtual} the virtual interval $\tau = \rho \cap
I$ is added to $V$ and it must still be there at time $t$ due to (P1).
So, assume that $\leftPoint(I)$ belongs to some virtual interval $\tau \in
V_{t - 1}$.
Since $\tau$ intersects with $\rho$, the inductive hypothesis on (P3)
implies that $\tau \subset \rho$ with a common endpoint.
In line~\ref{algline:ReplaceVirtual}, $\tau$ is replaced in $V$ by the new
virtual interval $\tau' = \tau \cap I$, which, by (P1) remains in $V$ at time
$t$.
The interval $\tau'$ intersects with both $\rho$ and $I$, hence, by
(P3) (applied to time $t$), it is contained in both of them, having a
common endpoint with each, thus $\tau' = \rho \cap \sigma$ and the property
holds.

\noindent
\textbf{Establishing (P5).}
Suppose toward a contradiction that there exists two distinct
intervals $\rho, \sigma \in V_{t}$ such that $\rho \cap \sigma \neq \emptyset$.
Assume without loss of generality that $\sigma$ was added to $V$ after
$\rho$.
By the inductive hypothesis, $\sigma$ is added to $V$ in round $t$, while (P2)
guarantees that $\rho \in V_{t - 1}$.
If $\sigma$ is added to $V$ in line~\ref{algline:ReplaceVirtual}, then
$\sigma = I \cap \tau$ for some virtual interval $\tau$ which is guaranteed to
be in $V_{t - 1}$ by (P2).
But then the inductive hypothesis implies that $\rho \cap \tau = \emptyset$,
thus $\rho \cap \sigma = \emptyset$.

So, assume that $\sigma$ is added to $V$ in
line~\ref{algline:AddNewVirtual}.
In that case $\sigma = I \cap \tau$ for some $\tau \in A_{t - 1}$.
Assume without loss of generality that $\leftPoint(\tau) < \leftPoint(I) <
\rightPoint(\tau) < \rightPoint(I)$ so that $\sigma = [\leftPoint(I),
\rightPoint(\tau)]$ is added to $V$ for $p = \leftPoint(I)$.
Since $\rho$ intersects with $\sigma$, it must also intersect with both $I$ and
$\tau$.
We know that $p$ cannot belong to $\rho$ as otherwise, the else
condition in line~\ref{algline:Else} would not have passed.
But, by the inductive hypothesis on (P3), $\rho \subset \tau$, thus $\rho
\subseteq I$ and $I$ should have been rejected in line~\ref{algline:Reject}.

Properties (P6) and (P7) can now be established based on the other
properties.

\noindent
\textbf{Establishing (P6).}
Consider some point $p \in \Reals$ and suppose toward a contradiction
that there exist three distinct intervals $\rho_1, \rho_2, \rho_3 \in A_{t}$
such that $p \in \rho_{i}$ for every $1 \leq i \leq 3$.
By (P4), the intersections $\sigma_{1, 2} = \rho_{1} \cap \rho_{2}$,
$\sigma_{1, 3} = \rho_1 \cap \rho_3$, and $\sigma_{2, 3} = \rho_2 \cap \rho_3$
are all in $V_{t}$.
But (P5) implies that $\sigma_{1, 2}$, $\sigma_{1, 3}$, and $\sigma_{2, 3}$
are pairwise disjoint, in contradiction to their definition.

\noindent
\textbf{Establishing (P7).}
Consider any two intervals $\rho, \sigma \in A_{t}$.
If $\rho \cap \sigma \neq \emptyset$, then (P4) implies that $\rho \cap
\sigma \in V_{t}$.
By (P3), $\rho \cap \sigma$ is strictly contained in both $\rho$ and $\sigma$,
hence $\rho$ cannot be a subset of $\sigma$ (nor can $\sigma$ be a subset of
$\rho$).
\end{proof}

%%%%%%%%%%%%%%%%%%%%%%%%%%%%%%%%%%%%%%%
\subsubsection{The components}
%%%%%%%%%%%%%%%%%%%%%%%%%%%%%%%%%%%%%%%
We employ Lemma~\ref{lemma:Structural} in order to understand the structure
of the components of $A$ and their relations with the intervals in
$V$.
To that end, fix some time $t$ and consider an arbitrary component
$C$ formed as the union of the actual intervals $\rho_1, \dots, \rho_k \in
A_{t}$.
We denote the leftmost and rightmost points in (the segment) $C$ by
$\leftPoint(C)$ and $\rightPoint(C)$, respectively.

\begin{figure}
\begin{center}
\includegraphics[width=0.6\linewidth]{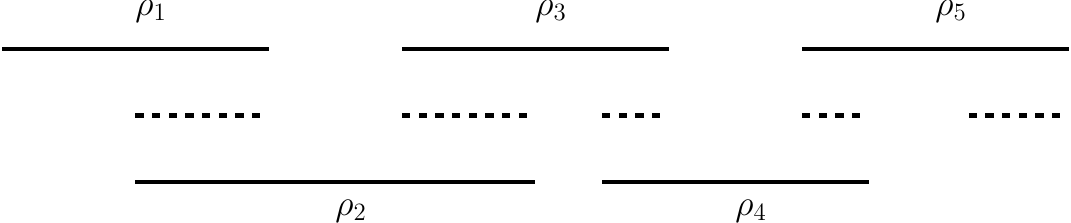}
\end{center}
\caption{\label{figure:Component}
A component $C$ of $A$.
The solid lines depict the actual interval $\rho_i$, $i = 1, \dots, 5$;
the dashed lines depict the virtual intervals that intersect with $C$ (all of
them must be contained in $C$).
}
\end{figure}

\sloppy
Assume without loss of generality that $\leftPoint(\rho_i) <
\leftPoint(\rho_{i + 1})$ for every $1 \leq i \leq k - 1$.
Lemma~\ref{lemma:Structural}(P6) and (P7) then guarantee that
\[
\leftPoint(\rho_{i - 1}) < \leftPoint(\rho_{i}) < \rightPoint(\rho_{i - 1}) <
\leftPoint(\rho_{i + 1}) < \rightPoint(\rho_i) < \rightPoint(\rho_{i + 1})
\]
for every $2 \leq i \leq k - 1$.
By Lemma~\ref{lemma:Structural}(P4), we conclude that $\rho_i \cap
\rho_{i + 1} \in V_{t}$ for every $1 \leq i \leq k - 1$, while
Lemma~\ref{lemma:Structural}(P3) implies that the segment $[\leftPoint(\rho_2),
\rightPoint(\rho_{k - 1})]$ does not intersect with any other virtual interval
in $V_{t}$.
The segment $C$ possibly contains two more virtual intervals at time $t$:
an interval $\sigma_{\ell} \subseteq [\leftPoint(\rho_1), \leftPoint(\rho_2))$
and an interval $\sigma_{r} \subseteq (\rightPoint(\rho_{k - 1}),
\rightPoint(\rho_k)]$, but then Lemma~\ref{lemma:Structural}(P3) guarantees
that $\leftPoint(\sigma_{\ell}) = \leftPoint(\rho_1) = \leftPoint(C)$ and
$\rightPoint(\sigma_{r}) = \rightPoint(\rho_k) = \rightPoint(C)$.
An illustration of a component is provided in Figure~\ref{figure:Component}.
There may also exist virtual intervals in between the components of
$A$, but we will soon show that their number and structure are fairly
limited.
This requires the definition of the following notions.
\par\fussy

%%%%%%%%%%%%%%%%%%%%%%%%%%%%%%%%%%%%%%%
\paragraph{Partition into portions.}
%%%%%%%%%%%%%%%%%%%%%%%%%%%%%%%%%%%%%%%
Point $p \in \Reals$ is said to be an \emph{$\Angles{x, y}$-point} at
time $t$ if
there exist exactly $x$ actual intervals in $A_{t}$ and $y$ virtual intervals
in $V_{t}$ that contain $p$.
By Lemma~\ref{lemma:Structural}, it suffices to consider $\Angles{x,
y}$-points for $\Angles{x, y}$ pairs in the set
$\mathcal{F} = \{ \Angles{0, 0}, \Angles{1, 0}, \Angles{0, 1}, \Angles{1, 1},
\Angles{2, 1} \}$,
referred to as \emph{feasible pairs}.
Fixing some feasible pair $\Angles{x, y} \in \mathcal{F}$,
a maximal connected set of $\Angles{x, y}$-points is referred to as an
\emph{$\Angles{x, y}$-portion}.
Given some segment 
$G \subseteq \Reals$,
let $\varphi(G)$ be the string of feasible pairs that encodes the types of
portions encountered in a left-to-right scan of $G$;
e.g., if $C$ is the component illustrated in Figure~\ref{figure:Component},
then
$\varphi(C) =
\Angles{1, 0} \,
\Angles{2, 1} \,
\Angles{1, 0} \,
\Angles{2, 1} \,
\Angles{1, 0} \,
\Angles{2, 1} \,
\Angles{1, 0} \,
\Angles{2, 1} \,
\Angles{1, 0} \,
\Angles{1, 1}$.

More generally, it follows from Lemma~\ref{lemma:Structural} and the discussion above that 
for every component $C$ of $A$ it holds that
\[
\varphi(C) \in
\Angles{1, 1}^{?} \, \Angles{1, 0} \, (\Angles{2, 1} \, \Angles{1, 0})^{*} \,
\Angles{1, 1}^{?} \, ,
\]
where following the common notation in regular expressions, we use a
superscript question mark to denote $0$ or $1$ occurrences and superscript
asterisk to denote zero or more occurrences (cf.\ the Kleene star).
Notice that by the distinct endpoints assumption, given $\varphi(C)$, it is
easy to determine (uniquely)
the numbers of actual and virtual intervals that $C$ contains and the
relative (total) order of their endpoints. Thus the string $\varphi(C)$
gives the so-called 
 {\em topological structure} of component $C$.
 
 Likewise, if $P$ is a segment between  two adjacent components of $A$, or
the segment to the left of the leftmost component of $A$, or the segment to
the right of the rightmost component of $A$, then it holds that 
\[
\varphi(P) \in
\Angles{0, 0} \, (\Angles{0, 1} \, \Angles{0, 0})^{*} \, .
\]
We refer to 
 the $\Angles{0, 1}$-portions as the  \emph{isolated} virtual
intervals.
The following lemma imposes a crucial restriction on the topological structure
of $\varphi(P)$  (and also  on $\varphi(C)$).

\begin{lemma} \label{lemma:zero-portions}
Let
$\varphi_{t} = \varphi(\Reals)$
be the string of feasible pairs that encodes the types of portions encountered
in a left-to-right scan of the real line at time $1 \leq t \leq n$.
Then, every $\Angles{0, 0}$ entry in $\varphi_{t}$ is immediately preceded or
immediately followed by a $\Angles{1, 0}$ entry.
\end{lemma}
\begin{proof}
By induction on $t$.
The assertion holds trivially for $t = 1$ as
$\varphi_{1} = \Angles{0, 0} \, \Angles{1, 0} \, \Angles{0, 0}$.
Suppose that the assertion holds at time $t - 1$ and consider time $t$.
If \Alg{} does not reach line~\ref{algline:AddNewActual} when processing
$I = I_{t}$,
then
$\varphi_{t} = \varphi_{t - 1}$
and the assertion holds by the inductive hypothesis, so assume hereafter that
\Alg{} does reach line~\ref{algline:AddNewActual} when processing $I$ which
means that
$I \nsupseteq J$
for any
$J \in A_{t - 1} \cup V_{t - 1}$.

The proof continues by case analysis that considers the different types of
possible intersections between $I$ and the portions corresponding to
$\varphi_{t - 1}$.
Specifically, in
Table~\ref{table:portion-intersection} we identify a total of 27 types of
intersections (up to symmetry) and address each one of them by depicting the
string $\varphi_{t - 1}$ together with a solid line stretched above the entries
corresponding to the portions with which interval $I$ intersects. The
resulting string $\varphi_{t}$ is given on the right column.
The strings $\alpha, \beta \in \mathcal{F}^{*}$ are used to denote
the substrings of $\varphi_{t - 1}$ encoding the portions in segments of
$\Reals$ which are (almost always) not affected by the newcoming interval $I$.
We also use the notation
\[
\alpha' =
\left\{
\begin{array}{ll}
\gamma, & \alpha = \gamma \, \Angles{0, 0} \\
\gamma \, \Angles{x - 1, y}, & \alpha = \gamma \, \Angles{x, y}, \, x > 0
\end{array}
\right.
\qquad
\beta' =
\left\{
\begin{array}{ll}
\gamma, & \beta = \Angles{0, 0} \, \gamma \\
\Angles{x - 1, y} \, \gamma, & \beta = \Angles{x, y} \, \gamma, \, x > 0
\end{array}
\right.
\]
which facilitates a reduction in the number of cases we have to consider.

Observe that $I$ can interest at most $2$ elements of $A\cup V$, since
otherwise $\exists J \in A \cup V$ s.t. $J \subseteq I$, and $I$ would not 
reach line~\ref{algline:AddNewActual} of the algorithm. Therefore, to cover all cases of the different types of
possible intersections between $I$ and the portions corresponding to
$\varphi_{t - 1}$, Table~\ref{table:portion-intersection} is divided into the
following 5 main categories:
\begin{itemize}

\item
C1:
$I$ intersects with $0$ components of $A$ and with $0$ isolated virtual
intervals.

\item
C2:
$I$ intersects with $0$ components of $A$ and with $1$ isolated virtual
interval.

\item
C3:
$I$ intersects with $1$ component of $A$ and with $0$ isolated
virtual intervals.

\item
C4:
$I$ intersects with $1$ component of $A$ and with $1$ isolated
virtual interval.

\item
C5:
$I$ intersects with $2$ components of $A$ and with $0$ isolated virtual
intervals.

\end{itemize}
For convenience,  Category C3 is further divided into two parts, where the bottom
part includes the cases where $I$ is included in the component  $C$ of $A$ that 
$I$ intersects (i.e $I - C = \phi$,)  and the top part includes the cases where $I - C \neq \phi$.

The assertion follows since in all cases, every $\Angles{0, 0}$ entry in
$\varphi_{t}$ is immediately preceded or immediately followed by a
$\Angles{1, 0}$ entry (taking into consideration the constraints imposed on
$\alpha$ and $\beta$ due to the inductive hypothesis).
\end{proof}

\begin{table}[!p]
\begin{center}
{\renewcommand{\arraystretch}{1.3}
\begin{tabular}{|c|c|c|}
\hline
category & $\varphi_{t - 1}$ & $\varphi_{t}$ \\
\hline
\multirow{1}{*}{C1}
&
$\alpha \,
\overline{\Angles{0, 0}} \,
\beta$
&
$\alpha \,
\Angles{0, 0} \,
\Angles{1, 0} \,
\Angles{0, 0} \,
\beta$
\\
\hline
\multirow{2}{*}{C2}
&
$\alpha \,
\overline{\Angles{0, 0} \, \Angles{0, 1}} \,
\Angles{0, 0} \,
\beta$
&
$\alpha \,
\Angles{0, 0} \,
\Angles{1, 0} \,
\Angles{1, 1} \,
\Angles{0, 0} \,
\beta$
\\
&
$\alpha \,
\Angles{0, 0} \,
\overline{\Angles{0, 1}} \,
\Angles{0, 0} \,
\beta$
&
$\alpha \,
\Angles{0, 0} \,
\Angles{1, 0} \,
\Angles{0, 0} \,
\beta$
\\
\hline
\multirow{13}{*}{C3}
&
$\alpha \,
\overline{\Angles{1, 0} \,
\Angles{0, 0}} \,
\beta$
&
$\alpha \,
\Angles{1, 0} \,
\Angles{2, 1} \,
\Angles{1, 0} \,
\Angles{0, 0} \,
\beta$
\\
&
$\alpha \,
\Angles{1, 0} \,
\overline{\Angles{1, 1} \,
\Angles{0, 0}} \,
\beta$
&
$\alpha \,
\Angles{1, 0} \,
\Angles{2, 1} \,
\Angles{1, 0} \,
\Angles{0, 0} \,
\beta$
\\
&
$\alpha \,
\Angles{0, 0} \,
\overline{\Angles{1, 1} \,
\Angles{1, 0} \,
\Angles{0, 0}} \,
\beta$
&
$\alpha \,
\Angles{0, 0} \,
\Angles{1, 1} \,
\Angles{1, 0} \,
\Angles{0, 0} \,
\beta$
\\
&
$\alpha \,
\Angles{1, 0} \,
\overline{\Angles{2, 1} \,
\Angles{1, 0} \,
\Angles{0, 0}} \,
\beta$
&
$\alpha \,
\Angles{0, 0} \,
\Angles{1, 0} \,
\Angles{2, 1} \,
\Angles{1, 0} \,
\Angles{0, 0} \,
\beta$
\\
& \ldots \ldots \ldots  \ldots \ldots \ldots \ldots \ldots \ldots \ldots \ldots \ldots \ldots \ldots 
           & \ldots \ldots \ldots \ldots \ldots \ldots \ldots \ldots \ldots \ldots \\
&
$\alpha \,
\overline{\Angles{1, 0}} \,
\beta$
&
$\alpha' \,
\Angles{0, 0} \,
\Angles{1, 0} \,
\Angles{0, 0} \,
\beta'$
\\
&
$\alpha \,
\Angles{1, 0} \,
\overline{\Angles{1, 1}} \,
\Angles{0, 0} \,
\beta$
&
$\alpha' \,
\Angles{0, 0} \,
\Angles{1, 0} \,
\Angles{0, 0} \,
\beta$
\\
&
$\alpha \,
\Angles{1, 0} \,
\overline{\Angles{2, 1}} \,
\Angles{1, 0} \,
\beta$
&
$\alpha' \,
\Angles{0, 0} \,
\Angles{1, 0} \,
\Angles{0, 0} \,
\beta'$
\\
&
$\alpha \,
\overline{\Angles{1, 0} \,
\Angles{1, 1}} \,
\Angles{0, 0} \,
\beta$
&
$\alpha' \,
\Angles{0, 0} \,
\Angles{1, 0} \,
\Angles{1, 1} \,
\Angles{0, 0} \,
\beta$
\\
&
$\alpha \,
\overline{\Angles{1, 0} \,
\Angles{2, 1}} \,
\Angles{1, 0} \,
\beta$
&
$\alpha' \,
\Angles{0, 0} \,
\Angles{1, 0} \,
\Angles{2, 1} \,
\Angles{1, 0} \,
\beta$
\\
&
$\alpha \,
\Angles{0, 0} \,
\overline{\Angles{1, 1} \,
\Angles{1, 0} \,
\Angles{1, 1}} \,
\Angles{0, 0} \,
\beta$
&
$\alpha \,
\Angles{0, 0} \,
\Angles{1, 1} \,
\Angles{1, 0} \,
\Angles{1, 1} \,
\Angles{0, 0} \,
\beta$
\\
&
$\alpha \,
\Angles{1, 0} \,
\overline{\Angles{2, 1} \,
\Angles{1, 0} \,
\Angles{1, 1}} \,
\Angles{0, 0} \,
\beta$
&
$\alpha \,
\Angles{1, 0} \,
\Angles{2, 1} \,
\Angles{1, 0} \,
\Angles{1, 1} \,
\Angles{0, 0} \,
\beta$
\\
&
$\alpha \,
\Angles{1, 0} \,
\overline{\Angles{2, 1} \,
\Angles{1, 0} \,
\Angles{2, 1}} \,
\Angles{1, 0} \,
\beta$
&
$\alpha \,
\Angles{1, 0} \,
\Angles{2, 1} \,
\Angles{1, 0} \,
\Angles{2, 1} \,
\Angles{1, 0} \,
\beta$
\\
\hline
\multirow{3}{*}{C4}
&
$\alpha \,
\overline{\Angles{1, 0} \,
\Angles{0, 0} \,
\Angles{0, 1}} \,
\Angles{0, 0} \,
\beta$
&
$\alpha \,
\Angles{1, 0} \,
\Angles{2, 1} \,
\Angles{1, 0} \,
\Angles{1, 1} \,
\Angles{0, 0} \,
\beta$
\\
&
$\alpha \,
\Angles{0, 0} \,
\overline{\Angles{1, 1} \,
\Angles{1, 0} \,
\Angles{0, 0} \,
\Angles{0, 1}} \,
\Angles{0, 0} \,
\beta$
&
$\alpha \,
\Angles{0, 0} \,
\Angles{1, 1} \,
\Angles{1, 0} \,
\Angles{1, 1} \,
\Angles{0, 0} \,
\beta$
\\
&
$\alpha \,
\Angles{1, 0} \,
\overline{\Angles{2, 1} \,
\Angles{1, 0} \,
\Angles{0, 0} \,
\Angles{0, 1}} \,
\Angles{0, 0} \,
\beta$
&
$\alpha \,
\Angles{1, 0} \,
\Angles{2, 1} \,
\Angles{1, 0} \,
\Angles{1, 1} \,
\Angles{0, 0} \,
\beta$
\\
\hline
\multirow{9}{*}{C5}
&
$\alpha \,
\overline{\Angles{1, 0} \,
\Angles{0, 0} \,
\Angles{1, 0}} \,
\beta$
&
$\alpha \,
\Angles{1, 0} \,
\Angles{2, 1} \,
\Angles{1, 0} \,
\Angles{2, 1} \,
\Angles{1, 0} \,
\beta$
\\
&
$\alpha \,
\Angles{1, 0} \,
\overline{\Angles{1, 1} \,
\Angles{0, 0} \,
\Angles{1, 0}} \,
\beta$
&
$\alpha \,
\Angles{1, 0} \,
\Angles{2, 1} \,
\Angles{1, 0} \,
\Angles{2, 1} \,
\Angles{1, 0} \,
\beta$
\\
&
$\alpha \,
\Angles{0, 0} \,
\overline{\Angles{1, 1} \,
\Angles{1, 0} \,
\Angles{0, 0} \,
\Angles{1, 0}} \,
\beta$
&
$\alpha \,
\Angles{0, 0} \,
\Angles{1, 1} \,
\Angles{1, 0} \,
\Angles{2, 1} \,
\Angles{1, 0} \,
\beta$
\\
&
$\alpha \,
\Angles{0, 0} \,
\overline{\Angles{1, 1} \,
\Angles{1, 0} \,
\Angles{0, 0} \,
\Angles{1, 1}} \,
\Angles{1, 0} \,
\beta$
&
$\alpha \,
\Angles{0, 0} \,
\Angles{1, 1} \,
\Angles{1, 0} \,
\Angles{2, 1} \,
\Angles{1, 0} \,
\beta$
\\
&
$\alpha \,
\Angles{0, 0} \,
\overline{\Angles{1, 1} \,
\Angles{1, 0} \,
\Angles{0, 0} \,
\Angles{1, 0} \,
\Angles{1, 1}} \,
\Angles{0, 0} \,
\beta$
&
$\alpha \,
\Angles{0, 0} \,
\Angles{1, 1} \,
\Angles{1, 0} \,
\Angles{1, 1} \,
\Angles{0, 0} \,
\beta$
\\
&
$\alpha \,
\Angles{1, 0} \,
\overline{\Angles{2, 1} \,
\Angles{1, 0} \,
\Angles{0, 0} \,
\Angles{1, 0}} \,
\beta$
&
$\alpha \,
\Angles{1, 0} \,
\Angles{2, 1} \,
\Angles{1, 0} \,
\Angles{2, 1} \,
\Angles{1, 0} \,
\beta$
\\
&
$\alpha \,
\Angles{1, 0} \,
\overline{\Angles{2, 1} \,
\Angles{1, 0} \,
\Angles{0, 0} \,
\Angles{1, 1}} \,
\Angles{1, 0} \,
\beta$
&
$\alpha \,
\Angles{1, 0} \,
\Angles{2, 1} \,
\Angles{1, 0} \,
\Angles{2, 1} \,
\Angles{1, 0} \,
\beta$
\\
&
$\alpha \,
\Angles{1, 0} \,
\overline{\Angles{2, 1} \,
\Angles{1, 0} \,
\Angles{0, 0} \,
\Angles{1, 0} \,
\Angles{1, 1}} \,
\Angles{0, 0} \,
\beta$
&
$\alpha \,
\Angles{1, 0} \,
\Angles{2, 1} \,
\Angles{1, 0} \,
\Angles{1, 1} \,
\Angles{0, 0} \,
\beta$
\\
&
$\alpha \,
\Angles{1, 0} \,
\overline{\Angles{2, 1} \,
\Angles{1, 0} \,
\Angles{0, 0} \,
\Angles{1, 0} \,
\Angles{2, 1}} \,
\Angles{1, 0} \,
\beta$
&
$\alpha \,
\Angles{1, 0} \,
\Angles{2, 1} \,
\Angles{1, 0} \,
\Angles{2, 1} \,
\Angles{1, 0} \,
\beta$
\\
\hline
\end{tabular}
}
\end{center}
\caption{\label{table:portion-intersection}
The different types of intersections of interval $I$ with the portions
corresponding to $\varphi_{t - 1}$ and the resulting $\varphi_{t}$.
}
\end{table}

%%%%%%%%%%%%%%%%%%%%%%%%%%%%%%%%%%%%%%%
\subsubsection{Accounting}
%%%%%%%%%%%%%%%%%%%%%%%%%%%%%%%%%%%%%%%
We are now ready to establish the following lemma.

\begin{lemma} \label{lemma:Approximation}
$|\Alg(S_{t})| \geq |\Opt(S_{t})| / 2$ at every time $0 \leq t \leq n$.
\end{lemma}
\begin{proof}
Lemma~\ref{lemma:Trace} guarantees that $|\Opt(S_{t})| \leq |\Opt(A_{t} \cup
V_{t})|$.
As $|\Alg(S_{t})| = |\Opt(A_{t})|$, it is sufficient to bound the ratio
$
R = \frac{|\Opt(A_{t} \cup V_{t})|}{|\Opt(A_{t})|}
$,
showing that it is at most $2$.

Let $C_{1}, \dots, C_{m}$ be the components of $A_{t}$ indexed from left to
right.
Lemma~\ref{lemma:zero-portions} implies that for every $1 \leq i \leq m - 1$,
at most one of the following three events occur:
(a) $\varphi(C_{i}) \in \mathcal{F}^{*} \, \Angles{1, 1}$;
(b) $\varphi(C_{i + 1}) \in \Angles{1, 1} \, \mathcal{F}^{*}$; or
(c) the segment $P$ between $C_{i}$ and $C_{i + 1}$ satisfies
$\varphi(P) = \Angles{0, 0} \, \Angles{0, 1} \, \Angles{0, 0}$ (i.e., there is
an isolated virtual interval between $C_{i}$ and $C_{i + 1}$).
Moreover, there is no isolated virtual interval to the left of $C_{1}$ or to
the right of $C_{m}$.
Clearly, the ratio $R$ can only increase if event (c) always occurs, so we
subsequently assume that this is indeed the case.
We will increase $R$ even further by assuming that there exists an isolated
virtual interval to the right of $C_{m}$.

Consider some component $C = C_{i}$, $1 \leq i \leq m$, and let
$A(C) = \{ \rho \in A_{t} \mid \rho \subseteq C \}$
and
$V(C) = \{ \sigma \in V_{t} \mid \sigma \subseteq C \}$.
It is easy to verify that $|V(C)| = |A(C)| - 1$ and that
$
|\Opt(A(C))| = \lceil |A(C)| / 2 \rceil
$,
whereas
$$
|\Opt(A(C) \cup V(C))|
~ = ~
\left\{
\begin{array}{ll}
|V(C)| & \text{ if } V(C) \neq \emptyset \\
1 & \text{ otherwise.}
\end{array}
\right.
$$
Accounting for the isolated virtual interval to the right of $C_{i}$, we
conclude that each component $C_{i}$, $1 \leq i \leq m$, contributes: \\
(i) $1$ to the denominator of $R$ and $2$ to the numerator of $R$, if $V(C_i) =
\emptyset$;
and \\
(ii) $\lceil |A(C_i)| / 2 \rceil$ to the denominator of $R$ and $|A(C_i)| - 1
+ 1 = |A(C_i)|$ to the numerator of $R$, if $V(C_i) \neq \emptyset$. \\
The assertion follows.
\end{proof}

\begin{corollary} \label{corollary:Approximation}
$|\Alg(S)| \geq |\Opt(S)| / 2$.
\end{corollary}

It remains to bound the space of our algorithm, showing that it is linear in
the length of the bit string representing $\Alg(S)$.
At each time $t$, the space of \Alg{} is linear in the length of
the bit strings representing $A_{t}$ and $V_{t}$.
As $\Opt(S_{t}) / 2 \leq \Alg(S_{t}) \leq \Opt(S_{t})$ for every $0 \leq t
\leq n$, and since $\Opt(S_{t})$ is non-decreasing with $t$, it is sufficient
to show that $|A_{t}| + |V_{t}| = O (|\Alg(S_{t})|) = O (|\Opt(A_{t})|)$.

By Lemma~\ref{lemma:Structural}(P6), we know that the actual intervals in
$A_{t}$ can be colored in two colors such that if two intervals belong to the
same color class, then they do not intersect.
Thus, $|A_{t}| \leq 2 \cdot |\Opt(A_{t})|$ at every time $t$.
On the other hand, Lemma~\ref{lemma:zero-portions} combined with our
understanding of the structure of the connected components imply that if we
count the actual and virtual intervals by scanning the real line from left to
right, then the number of virtual intervals never exceeds that of the actual
intervals (this is also showed in the proof of
Lemma~\ref{lemma:Approximation}).
Therefore, $|V_{t}| \leq |A_{t}|$ which establishes the following corollary.

\begin{corollary} \label{corollary:Space}
At every time $t$, the space of \Alg{} is linear in the length of the
bit string representing $\Alg(S)$.
\end{corollary}

\section{Proper Intervals}
\label{section:ProperIntervals}
%%%%%%%%%%%%%%%%%%%%%%%%%%%%%%%%%%%%%%%%%%%%%%%%%%%%%%%%%%%%%%%%%%%%%%%%%%%%%%

In this section we consider the interval selection problem for proper intervals.  There is an easy
deterministic $2$-approximate streaming (and online) algorithm that uses no extra space in addition to storing the
output: simply greedily add an interval whenever possible.  We give here a streaming algorithm with an improved
approximation ratio of $3/2$ that uses output-linear space. As we show in \Section{}~\ref{section:LowerBounds}, that is optimal.

% Information description
We first give an informal overview of the operation of the algorithm.  It maintains a collection of disjoint segments,
called \emph{zones}, which partitions the subset of the real line covered by the intervals seen so far.  For each zone,
the algorithm keeps track of two intervals from the input stream: the one with the leftmost left endpoint (reps., rightmost
right endpoint) among those with a left (resp., right) endpoint in the zone.  The output of the algorithm is the
maximum interval selection from this set of recorded intervals. The performance guarantee essentially follows from
showing that this solution contains at least two intervals within any span of three intervals of the optimal solution. 

% Defining zones
We now define the zones.
Define the \emph{support} $supp(X)$ of a set $X$ of intervals to be the subset of the real line covered by all intervals in
$X$, or $supp(X) = \cup_{I \in X} I$.  
% A \emph{component} is a maximal continuous segment formed by a union of zones.
{\em Connected components} of the support of $X$ are defined in a natural manner. 
The maximal segments on the real line that are outside connected components of the support of
$X$ are called \emph{out-regions} (with respect to $X$).
The zones will be  segments on the real line and the 
zone collection $Z_t$ at time $t$ will be  a partition of $supp(S_t)$.  
A zone may initially be 
%Once created by the algorithm, a zone never disappears, but, at any time,  it can be 
\emph{flexible} in that its endpoints might change,
% (to extend its reach by merging it into an adjacent zone), 
%and
and  in that it may be merged with other regions or zones into a single zone.
%% Other {\em fixed}.
Zones at the extreme ends of connected components are flexible, with the sole exception 
of the initial zone of a connected component;
other zones are \emph{fixed} and unchangeable, with permanent endpoints.

% Maintaining the zones
 We now specify how the algorithm  creates and maintains the zones.
 Initially, there are no zones and the whole line is considered one out-region.
 When an interval $I$ is received at time $t$, we consider the following cases depending on the positions of its endpoints:

\begin{enumerate}

\item \emph{Both endpoints of $I$ are in an out-region:} Create a new fixed zone defined by the endpoints of  $I$, in a connected component of its own.
%$[\lefty(I), \righty(I))$ and a new degenerate flexible zone $[\righty(I),\righty(I)]$.

\item \emph{Both endpoints of $I$ fall within the same connected component:} Do nothing.

\item \emph{The endpoints of $I$ belong to zones in different connected components:} First,  for each of the 
two zones in which the endpoints fall, fix it, if it is not already fixed, without changing its endpoints.
Then, create a new fixed zone which includes the out-region 
  between the respective components; denote it $z$. If $I$ properly includes one or two  flexible 
  zones, then these zones
  are merged into $z$; the resulting zone  $z$ is
a fixed zone with right (reps., left) endpoint changed to be the right (resp., left) endpoint
of the flexible zone to the right (reps., left) of $z$.
%  %Observe that each such flexible zone, if it exists, is merged into the new zone. 
%  That is, if
%$I$ properly includes flexible  zones at the extremes of two components, these flexible zones are merged into the new fixed
%zone created from the out region.

\item \emph{One endpoint of $I$ falls in a zone and the other in an out-region:} 
Let $k$ be the zone, and $C$ be the connected component, in which one endpoint of $I$ falls.
Without loss of generality assume that  the right endpoint of $I$ falls in the out-region, i.e., the left endpoint of $I$ falls in $k$.
Fix zone $k$ if it is not yet fixed, without changing its endpoints.
Create a new flexible zone which covers $I \setminus supp(C)$; denote it $z$.
If $I$ properly includes a flexible zone, denote it $k'$,  then $k'$ is merged into $z$, where
the resulting  zone $z$  has as its left endpoint the left endpoint $k'$. 
  \end{enumerate}

\begin{figure}
\begin{center}
\includegraphics[width=0.75\linewidth]{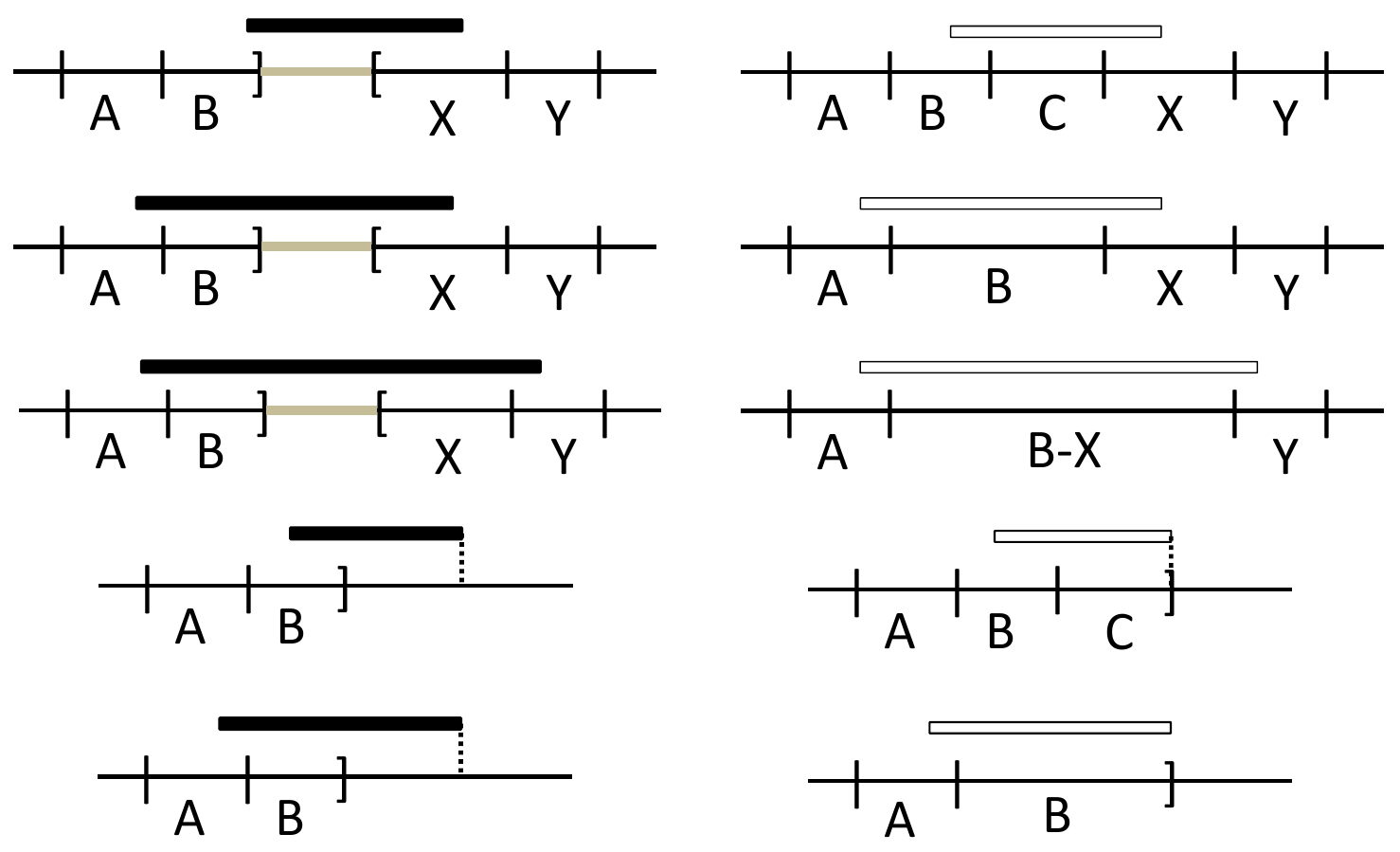}
\end{center}
\caption{\label{fig:zones} Illustration of the different cases in zone management, with three variations of Case 3 and two for Case 4. On the left, an incoming interval is shown in bold with the prior settings of the zones; the resulting zone configuration is shown on the right. Flexible ends of zones are shown with brackets and out-regions in grey.}
\end{figure}

This completes the specification of the zones. See Fig.~\ref{fig:zones} for illustration of the different cases.

It is  clear from the definition of the zones and their endpoints  that the zone collection partitions the support
of the intervals seen so far.
We shall assume a total order on the zones, from left to right.
The following lemma is crucial for the correctness and performance guarantee of the algorithm.

%% Invariants
\begin{lemma}\label{lem:properties}
The zone collection satisfies the following invariants:
%\begin{quote}
\begin{description}
  \item[(\#)] Each zone is properly contained in an input interval, and
%  \item[(*)] Every endpoint of a zone is an endpoint of an interval, and
% \item[$\dagger$] A flexible zone is defined by an interval with its other endpoint in an adjacent zone.
  \item[(*)] Each input interval properly contains at most one zone, except
   those intervals that  when they arrive fall under Case 3 of the algorithm, which can contain two zones.
\end{description}
%\end{quote}
\end{lemma}

\begin{proof}
The former invariant follows directly from the observation that
in each case that a zone is created 
%or is extended 
(cases 1, 3 and 4), 
the zone is properly contained in the input interval.

To prove the latter invariant by \emph{reductio ad absurdum}, 
assume that there is an input sequence for which the invariant fails.
Let time $t$ be the first time when the invariant does not hold 
on this input. Let $I_{t''}$, $t''\leq t$ be interval that contains three
zones, and let these zones be (from left to right)
$k_1, k_2$, and $k_3$.
Let $t'$ be the time when zone $k_2$ achieved its extent as it is at time $t$. 

We consider the four cases that can  govern the treatment of $I_{t'}$ by the algorithm.

In Case 1,  the only zone that is created or changes its endpoint is the zone with endpoints equal
to those of $I_{t'}$. It follows that  $I_{t'} \subset I_{t''}$,
which contradicts the assumption that the input  intervals are proper intervals.

In Case 2, no zones are created or change their endpoints.

In Case 3, the only zone that is created or changes its endpoints is the zone $z$ that is
created and includes the out-region contained in the input interval $I_{t'}$. Furthermore, 
at the end of time $t'$, zone $z$ has one adjacent zone to its right, and one
adjacent zone to its left, which include the right endpoint and left endpoint or $I_{t'}$, 
respectively. These zone must thus be zones $k_1$ and $k_2$. It follows that 
 $I_{t'} \subset I_{t''}$, a contradiction to the assumption that the input  intervals are proper intervals.

Finally, consider Case 4.
Without loss of generality assume that the  right endpoint  of $I_{t'}$ fell in the out-region.
The only zone that is created or changes its endpoint in this case is zone $z$,  which has, at the end of time
$t'$, its right endpoint equal to the right endpoint of $I_{t'}$. Furthermore, the left endpoint of
$I_{t'}$ falls in the zone adjacent to $z$ and to the left of $z$, i.e., in zone $k_1$.  It follows again  
that  $I_{t'} \subset I_{t''}$,
which contradicts the assumption that the input  intervals are proper intervals.
%% Thus, the zone collection $Z_t$ satisfies the invariant.
%Zone $k_2$ was an extreme zone of a component in $Z_{t'}$.
%Let us suppose without loss of generality that it was the rightmost point of the component.
%We claim that the left endpoint of $I_{t'}$ fell in an adjacent zone, i.e., $k_1$.
%That would imply that $I_{t'}$ is properly included in $I_{t+1}$, which is a contradiction.
%
%To establish the claim, consider the two sub-cases that occur in Case 4.  If $\lefty(I_{t'})$ fell in a
%flexible zone, or one that was previously extreme in the component (i.e., an initial zone), then that zone is $k_1$, and
%the claim follows.  Otherwise, $\righty(I_{t'})$ extended a flexible zone; let $J$ denote the interval that
%previously supplied its extreme endpoint.  Since the invariant held when $J$ was introduced, $\lefty(J)$
%was in the adjacent zone, namely $k_1$.  Thus, $I_{t'}$ cannot contain zone $k_1$, as it would then contain $J$.  Thus,
%$\lefty(I_{t'})$ falls in zone $k_1$.  Hence, the claim and the lemma.
\end{proof}

% \begin{proof}
% We claim that every six contiguous zones properly contain two disjoint input intervals.
% They must therefore properly contain at least one interval from $OPT$.
% To establish the claim, number the zones from 1 through 6.
% The interval that contains zone 2 (zone 5) must be properly contained.

% The intervals maintained by the algorithm
We next define the intervals maintained by the algorithm.  For an interval $I$ and time $t$, let $\bzone(I) = \bzone_t(I)$
(resp. $\fzone(I) = \fzone_t(I)$) be the zone in which $\lefty(I)$ (resp. $\righty(I)$) falls at time $t$.
%%% define $D_t = \{\bzone(I),\fzone(I) : I \in S_t\}$ to be the set of zones that include at least one endpoint of an interval in $S_t$.  
For each zone $k$,  and time $t$, the algorithm keeps  the endpoints of the intervals $L_k = \argmin_{I \in S_t, \bzone(I)=k, } \lefty(I)$, i.e., the
one with the leftmost left endpoint in the zone, and $R_k = \argmax_{I\in S_t, \fzone(I)=k} \righty(I)$, i.e., the one with
the rightmost right endpoint in the zone. Note that $L_k$ (resp., $R_k$) will be undefined if no interval has its
left (resp., right) endpoint in zone $k$.
% For convenience we write for a zone $k$ that contains no endpoint of an interval that $L_k =R_{k'}$, where
% $k'=\max_q \{q \in D_t : q < k\}$, and $R_k = L_{k''}$, where $k''=\min_q \{q \in D_t : q > k\}$.

The output of the algorithm is the maximum interval selection of the set
$A = \cup_k \{L_k, R_k\}$, obtained by, e.g.,  the classic left-right greedy algorithm.
%This completes the specification of the algorithm.

%Invariants:
%(2)  each component of $L_t$ will include a closed segment.
%(2) Each zone is properly contained in an input interval.

We next detail how the algorithm maintains the intervals stored.
When interval $I$ arrives at time $t$, we consider the four cases defined above.

If Case 1 applies, then $I$ defines a new zone $k$,
and both $L_k$ and $R_k$ are permanently set to equal $I$.

If Case 2 applies, no zones change;
%the endpoints of $I$ fall in zones fixed zones, or zones that become fixed in that round,
the $L$- and $R$-intervals are updated for the zones in which the endpoints of $I$ fall.
% In Cases 2, 3 and 4, where an endpoint of $I$ falls in a fixed zone, or a zone that is fixed
% in that round, then $L_k$ or $R_k$ are updated as appropriate.
E.g., if $b(I)$ falls in zone $k$, and $b(I) < b(L_k)$, then $L_k$ is updated as $I$.

If Case 3 applies,  the $L$- and $R$-intervals are updated, for the zones in which the endpoints of $I$ fall.
Then, a new zone is created, but all the endpoints, if any, that fall in that zone have
already appeared, and thus we need only merge the information from the at most two
previous flexible zones that are merged into the new zone.

If Case 4 applies, then one endpoint of $I$ falls in a zone $k$, which becomes fixed if
it was not fixed before.
Without loss of generality, assume it is the left endpoint of $I$ that falls in $k$.
So, $L_k$ is updated appropriately.
Then, a new zone is created; denote it $z$. If $I$ does not properly include any flexible zone, then $z$ 
covers $I \setminus supp(C)$,  $R_z$ is defined to be $I$, and $L_z$ is undefined.
If $I$ does properly  include a flexible zone, denote to $k'$, then $k'$ is merged into $z$.
$R_z$ is defined to be $I$, and $L_z$ is defined to be $L_{k'}$.

% In one sub-case, the right endpoint falls in a new zone $k'$,
%whose only endpoint is $\righty(I)$; then, $R_{k'}$ is set as $I$ and $L_{k'}$ is undefined.
%In the other sub-case, the right endpoint falls in a flexible zone $k'$ (at the edge of a component)
%that is extended to include $\righty(I)$. In this case, $\righty(I)$ is also the right endpoint of zone $k'$,
%so it is clearly also the rightmost right endpoint that falls in the zone.
%No left endpoint falls in zone $k'$ in this round, so $L_{k'}$ stays unchanged (and undefined).

We conclude that the algorithm can indeed maintain the intervals $L_k$ and $R_k$ for each zone $k$ as defined.

\smallskip

An immediate corollary of Lemma~\ref{lem:properties} is that the union of any five adjacent zones must properly contain an input
interval. By induction, any $5t$ adjacent zones must include at least $t$ disjoint intervals.
% The union of six adjacent zones must then properly contain an interval from the optimal solution $OPT$.
Hence, we obtain an upper bound on the space used by the algorithm  by noting 
that for each zone $k$, only six values need to be recorded: the endpoints of $L_k$, of $R_k$ and of the zone $k$ itself. We thus have the following.
\begin{lemma}
The number of zones is at most $5 |OPT|+4$.
The space used by the algorithm is at most $O(|OPT|)$.
%The amount of information stored is then $O(|OPT|) = O(|A|)$.
\end{lemma}

Finally, we prove the performance guarantee of the algorithm.
The following lemma captures the core of the argument.
For two proper intervals $I$ and $J$, we write $I \le J$ to denote that $\lefty(I) \le \lefty(J)$ (and thus also
$\righty(I)\le \righty(J)$), and similarly $I < J$ if $\lefty(I) < \lefty(J)$.

 \begin{lemma}
Let $R \subseteq S_t$ be a collection of three disjoint input intervals.
Then, at the end of round $t$, $A$ contains a pair of intervals  of $S_t$ that are 
contained in the span of $R$.
\label{lem:triple-span}
\end{lemma}

\begin{proof}
Let the three disjoint intervals be $O_1 < O_2 < O_3$.
Our claim is that $A$ contains a pair of disjoint intervals $I, I'  \in S_t$, $I, I' \subset
[\lefty(O_1), \righty(O_3))$.

Let $b_i = \bzone(O_i)$ ($f_i = \fzone(O_i)$) be the zone in which the left (right) endpoint of $O_i$ falls, for $i=1,2,3$, respectively. 
Clearly, $b_1 \le f_1 \le b_2 \le f_2 \le b_3 \le f_3$.
We observe that $b_1 < b_2 < b_3$.
Suppose, e.g., that $b_1 = b_2$. Then, by 
Lemma~\ref{lem:properties}(Invariant (\#)), there is an interval that properly contains $O_1$, a contradiction.
%Therefore, if $f_1 = \fzone(O_i)$ is the zone in which the right endpoint of $O_i$ falls, 
Therefore, $f_1 < b_3$.
%and $f_i = \fzone(O_i)$, 
% and note that by Lemma~\ref{lem:properties}(Invariant (\#)),
% $b_1 < f_1 \le b_2 < f_2 \le b_3 < f_3$.

Consider $R_{f_1}$ and $L_{b_3}$, which are well-defined intervals since the right endpoint of 
$O_1$ falls in zone $f_1$ and the left endpoint of $O_3$ falls in zone $b_3$.
By definition, $O_1 \le R_{f_1}$ and $L_{b_3} \le O_3$, so $R_{f_1}, L_{b_3} \subset [\lefty(O_1),\righty(O_3))$.  
Since $f_1 < b_3$, it follows that $R_{f_1}$ and $L_{b_3}$ are disjoint, establishing the
claim.
\end{proof}

We conclude with the following theorem.

\begin{theorem}
$|ALG| \ge 2|OPT|/3$.
\end{theorem}

\begin{proof}
Let $|S|=n$, and 
let $O_0, O_1, \ldots, O_{p-1}$ be the intervals in the optimal
solution in order of (left) endpoints, where $p = |OPT|$.

If $p \pmod{3} \equiv 0$, we have by Lemma \ref{lem:triple-span}
that the algorithm obtains at least two intervals within each of  the 
segments $[\lefty(O_{3r}), \righty(O_{3r+2)}))$, for  $ 0 \leq r \leq (p-3)/3$.
Thus, we have that $|ALG| \ge 2((p-3)/3 + 1) = 2p/3$.

If $p \pmod{3} \equiv 1$ we have by Lemma \ref{lem:triple-span}
that the algorithm obtains at least two intervals within each of  the 
segments $[\lefty(O_{3r}), \righty(O_{3r+2)}))$, for  $ 0 \leq r \leq (p-4)/3$.
Now, if $k_{max}$  is the last zone at the end of 
round $n$ then $O_{p-1} \le R_{k_{max}}$; thus the 
algorithm obtains an interval that intersects only  $O_{p-1}$.
Together we get that $|ALG| \ge 2((p-4)/3+1)+1 > 2p/3$.

If $p \pmod{3} \equiv 2$,  we have by Lemma \ref{lem:triple-span}
that the algorithm obtains at least two intervals within each of  the 
segments $[\lefty(O_{3r+1}), \righty(O_{3r+3)}))$, for  $ 0 \leq r \leq (p-5)/3$.
Now, if $k_{min}$ and $k_{max}$ are the first and last 
zones at the end of round $n$ then, $L_{k_{min}} \le O_0$ and $O_{p-1} \le R_{k_{max}}$;
thus the algorithm obtains an interval that intersects only $O_0$
and another that intersects only $O_{p-1}$.
Together we get that $|ALG| \ge 2 ((p-5)/3+1)+2 > 2p/3$.
\end{proof}

%%%%%%%%%%%%%%%%%%%%%%%%%%%%%%%%%%%%%%%%%%%%%%%%%%%%%%%%%%%%%%%%%%%%%%%%%%%%%%
\section{Lower Bounds}
\label{section:LowerBounds}
%%%%%%%%%%%%%%%%%%%%%%%%%%%%%%%%%%%%%%%%%%%%%%%%%%%%%%%%%%%%%%%%%%%%%%%%%%%%%%
In this section we establish lower bounds on the approximation ratio of
randomized streaming algorithms for the interval selection problem,
establishing the following two theorems.

\begin{theorem}[Lower bound for general intervals]
\label{theorem:LowerBoundGeneral}
For every real $\epsilon > 0$, integers $k_0, n_0 > 0$, and subexponential
(respectively, sublinear) function $s : \Naturals \rightarrow \Naturals$,
there exist $k_0 \leq k \leq c \cdot k_0$, where $c$ is a universal constant,
$n > n_0$, and an interval stream $S$ such that \\
(1) $|S| = n$; \\
(2) $|\Opt(S)| = k$; and \\
(3) $\Alg(S) < k (1 / 2 + \epsilon)$ for any randomized interval selection
streaming algorithm \Alg{} with space $s(k b)$ (resp., space $s(n b)$), where
$b$ is the length of the bit strings representing the endpoints.
\end{theorem}

\begin{theorem}[Lower bound for unit intervals]
\label{theorem:LowerBoundProper}
For every real $\epsilon > 0$, integers $k, n_0 > 0$, and subexponential
(respectively, sublinear) function $s : \Naturals \rightarrow \Naturals$,
there exist $n > n_0$, and a unit interval stream $S$ such that \\
(1) $|S| = n$; \\
(2) $|\Opt(S)| = k$; and \\
(3) $\Alg(S) < k (2 / 3 + \epsilon)$ for any randomized proper interval
selection streaming algorithm \Alg{} with space $s(k b)$ (resp., space $s(n
b)$), where $b$ is the length of the bit strings representing the endpoints.
\end{theorem}

Our lower bounds are proved by designing a random interval stream
$S$ for which every deterministic algorithm performs badly on expectation;
the assertion then follows by Yao's principle.
(Our construction uses half-open intervals, but this can be easily altered.)
Note that under the setting used by our lower bounds, the algorithm is
required to output a collection $\mathcal{C}$ of disjoint intervals, and the
quality of the solution is then determined to be the cardinality of
$\mathcal{C} \cap S$.
In other words, the algorithm is allowed to output non-existing intervals (that
is, intervals that never arrived in the input), but it will not be credited
for them.
This, obviously, can only increase the power of the algorithm.

%%%%%%%%%%%%%%%%%%%%%%%%%%%%%%%%%%%%%%%
\paragraph{The $(k, n)$-gadget.}
%%%%%%%%%%%%%%%%%%%%%%%%%%%%%%%%%%%%%%%
Fix some positive integer $m$ whose role is to bound the space of the
algorithm.
Our lower bounds rely on the following framework, characterized by the
parameters $k, n \in \Integres_{> 0}$, denoted a \emph{$(k, n)$-gadget}.
Consider an extensive form two-player zero-sum game played between the
algorithm (MAX) and the adversary (MIN), depicted by a sequence of $k$
\emph{phases}.
Informally, in each phase $t$, the adversary chooses a permutation $\pi_{t}
\in P_n$, where $P_n$ is the collection of all permutations on $n$ elements,
and an index $i_t \in [n]$.
The algorithm observes $\pi_t$ (but not $i_t$) and produces a \emph{memory
image} $M_t$, i.e., a bit string of length $m$.
The index $i_t$ is handed to the algorithm after the memory image is
produced.
At the end of the last phase the algorithm tries to \emph{recover}
$\pi_t(i_t)$ for $t = 1, \dots, k$:
it outputs some $i_{t}^{*} \in [n]$ based on the memory image $M_t$, index $i_t$,
and all other memory images and indices.
For each $t$ such that $i_{t}^{*} = \pi_t(i_t)$, the algorithm scores a
(positive) point.

More formally, the adversarial strategy is depicted by the choices of
the permutations $\pi_t$ and the indices $i_t$ for $t = 1, \dots, k$.
We commit the adversary to make those choices uniformly at random (so, the
adversary reveals its mixed strategy), namely, $\pi_t \in_{r} P_n$ and $i_t
\in_{r} [n]$ for every $t$, where all the random choices are
independent.
The strategy of the algorithm is depicted by the function sequences $\{ f_t
\}_{t = 1}^{k}$ and $\{ g_t \}_{t = 1}^{k}$, where
$$
f_t : P_n \times \left( \{0, 1\}^m \times [n] \right)^{t - 1} \rightarrow \{0,
1\}^m
\qquad \text{and} \qquad
g_t : \{0, 1\}^m \times [n] \times \left( \{0, 1\}^m \times [n] \right)^{k -
1} \rightarrow [n] ~ .
$$
Let $\Gamma_0$ be the empty string and recursively define\footnote{
We use the notation $u \circ v$ to denote the concatenation of the string $u$
to string $v$.
}
$
\Gamma_t = \Gamma_{t - 1} \circ f_t \left( \pi_t, \Gamma_{t - 1} \right) \circ i_t
$.
The payoff of the algorithm is the number of $t$s, $1 \leq t \leq k$, such
that
$$
g_t \left( f_t \left( \pi_t, \Gamma_{t - 1} \right), i_t, \left\{ f_{t'}
\left( \pi_{t'}, \Gamma_{t' - 1} \right), i_{t'} \right\}_{t' \neq t} \right)
~ = ~
\pi_t(i_t) ~ .
$$

In the language of the aforementioned informal description, the role of the
function $f_t$ is to produce the memory image $M_t$ based on the permutation
$\pi_t$ and all previous memory images and indices (whose concatenation is
given by $\Gamma_{t - 1}$).
The role of the function $g_t$ is to recover $\pi_t(i_t)$ based on the memory
image $M_t$, index $i_t$, and all other memory images and indices.

Note that the memory images $M_{t'}$ and indices $i_{t'}$, $t' \neq t$, do not
contain any information on the permutation $\pi_t$ on top of that contained in
$M_t$.
In particular, the entropy in $\pi_{t}(i_t)$ given $M_t$, $i_t$, and $\{ M_{t'},
i_{t'} \}_{t' \neq t}$ is equal to the entropy in $\pi_{t}(i_t)$ given $M_t$
and $i_t$.
Therefore, it will be convenient to decompose the domain of the function $g_t
: \{0, 1\}^m \times [n] \times (\{0, 1\}^m \times [n])^{k - 1} \rightarrow
[n]$ so that the $(\{0, 1\}^m \times [n])^{k - 1}$-part determines which
function $\hat{g}_t : \{0, 1\}^m \times [n] \rightarrow [n]$ is chosen, and then
this function $\hat{g}_t$ is used to produce $i_{t}^{*}$ based on $M_t$ and
$i_t$.
Similarly, we decompose the domain of the function $f_t : P_n \times (\{0,
1\}^m \times [n])^{t - 1} \rightarrow \{0, 1\}^m$ so that the $\left( \{0,
1\}^m \times [n] \right)^{t - 1}$-part determines which function $\hat{f}_t :
P_n \rightarrow \{0, 1\}^m$ is chosen, and then this function $\hat{f}_t$ is
used to produce $M_t$ based on $\pi_t$.

We now turn to bound the expected payoff of the algorithm as a function of
$k$, $m$, and $n$.
The key ingredient in this context is the following lemma, which is
essentially a well known fact in slightly different settings;
a proof is provided in
Appendix~\ref{section:ProofOfLemmaImpossibleCode}
for completeness.

\begin{lemma} \label{lemma:ImpossibleCode}
For every real $\alpha > 0$ and integer $n_0 > 0$, there exists an integer $n
> n_0$ such that for every two functions $\hat{f} : P_n \rightarrow \{0,
1\}^m$ and $\hat{g} : \{0, 1\}^m \times [n] \rightarrow [n]$, where $m =
\alpha n \log n$, we have
$\Probability_{\pi \in_{r} P_n, i \in_{r} [n]} \left( \hat{g}(\hat{f}(\pi), i)
= \pi(i) \right) < 2 \alpha$.
\end{lemma}

\begin{corollary} \label{corollary:Gadget}
For every real $\alpha > 0$ and integers $k, n_0 > 0$, there exists an integer
$n > n_0$ such that if $m \leq \alpha n \log n$, then the expected payoff of
the algorithm player in a $(k, n)$-gadget is smaller than $2 \alpha k$.
\end{corollary}

\begin{figure}
\begin{center}
\includegraphics[width=0.7\linewidth]{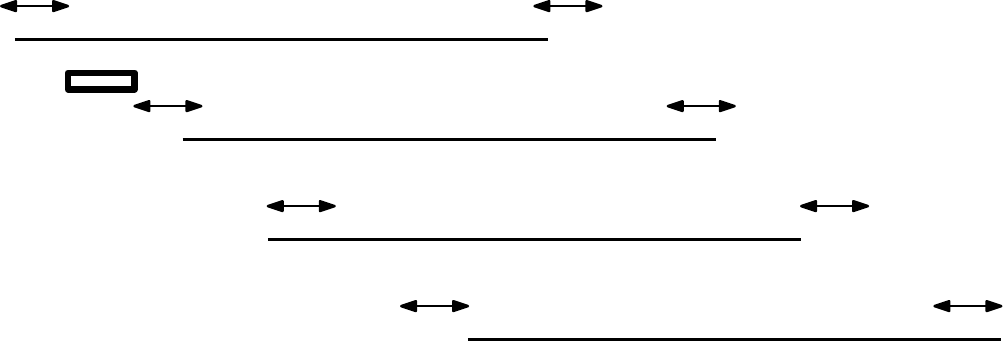}
\end{center}
\caption{\label{figure:Stack}
The relative locations of the intervals in an $(n, \pi)$-stack for $n = 4$.
The left and right endpoints of interval $J_i$ are located in the segments
depicted by the bidirectional arrows whose length is $\lambda / 2$.
The exact location within this segment is determined by $\pi(i)$.
In the construction of the $2$-lower bound for general intervals, the bold
rectangles correspond to the segments $\sigma_{\ell}$ and $\sigma_{r}$ in
which the stacks (or auxiliary intervals) identified with the left and right
children, respectively, of the current node are deployed assuming that the
good interval is interval $J_2$ (these segments do not intersect with the
segments corresponding to the bidirectional arrows).
}
\end{figure}

\sloppy
%%%%%%%%%%%%%%%%%%%%%%%%%%%%%%%%%%%%%%%
\paragraph{The $(n, \pi)$-stack.}
%%%%%%%%%%%%%%%%%%%%%%%%%%%%%%%%%%%%%%%
We now turn to implement a $(k, n)$-gadget via a carefully designed interval
stream.
As a first step, we introduce the \emph{$(n, \pi)$-stack} construction.
Given an integer $n > 0$ and a permutation $\pi \in P_n$, an $(n, \pi)$-stack
deployed in the segment $[x, y)$, $x < y$, is a collection of $n$ intervals
$J_1, \dots, J_n$
satisfying: \\
(1) all intervals $J_i$ are half open; \\
(2) all intervals $J_i$ have the same length $\rightPoint(J_i) -
\leftPoint(J_i) = \lambda n$, where $\lambda = \frac{y - x}{2 n - 1 / 2}$; and
\\
(3) $\leftPoint(J_i) = x + \lambda (i - 1) + \epsilon \pi(i)$ for every $i \in
[n]$, where $\epsilon = \lambda / (2 n)$. \\
Note that this deployment ensures that $\leftPoint(J_n) <
\rightPoint(J_1)$, hence the half open segment $[\leftPoint(J_n),
\rightPoint(J_1))$ is contained in $J_{i}$ for every $i \in [n]$.
Moreover, the union of the intervals in the stack does not necessarily cover
the whole segment $[x, y)$;
it is always contained in $[x, y)$, though.
The structure of an $(n, \pi)$-stack is illustrated in
Figure~\ref{figure:Stack}.
\par\fussy

The $(k, n)$-gadget is implemented by introducing $k$ stacks, each
corresponding to one phase, and some \emph{auxiliary} intervals;
the stack corresponding to phase $t$ is referred to as stack $t$.
The permutation $\pi$ used in the construction of stack $t$ is $\pi_t$.
The index $i_t$ will dictate the choice of one \emph{good} interval out of the
$n$ intervals in that stack.
What exactly makes this interval good will be clarified soon;
informally, the algorithm has no incentive to output an interval in a stack
unless this interval is good.

The $k$ stacks are used both by the construction of the $2$-lower bound
for general interval streams and by that of the $(3 / 2)$-lower bound for
unit intervals.
The difference between the two constructions lies in the manner in which
these stacks are deployed in the real line, and in the addition of the
auxiliary intervals.

%%%%%%%%%%%%%%%%%%%%%%%%%%%%%%%%%%%%%%%
\paragraph{A $(3 / 2)$-lower bound for unit intervals.}
%%%%%%%%%%%%%%%%%%%%%%%%%%%%%%%%%%%%%%%
The interval stream that realizes the $(k, n)$-gadget for the $(3 /
2)$-lower bound for unit intervals is constructed as follows.
It contains $k$ sufficiently spaced apart stacks, where the intervals in each
stack are scaled to a unit length (so $\lambda = 1 / n$).
Consider stack $t$ and suppose that it is deployed in the segment $[x, y)$,
where $y = x + 2 - 1 / (2 n)$.
Recall that the permutation that determines the exact location of the
intervals in the stack is $\pi_t$ and that the good interval is $J_{i_t}$.

After the arrival of the $n$ intervals in the stack, two more half open unit
auxiliary intervals are presented:
$$
L_t
~ = ~
\left[ x + \frac{i_t - 1}{n} - 1, x + \frac{i_t - 1}{n} \right)
\qquad \text{and} \qquad
R_t
~ = ~
\left[ x + \frac{i_t - 1 / 2}{n} + 1, x + \frac{i_t - 1 / 2}{n} + 2 \right) ~ .
$$
In other words, the interval $L_t$ (respectively, $R_t$) is located to the
left (resp., right) of the leftmost (resp., rightmost) point in which
$\leftPoint(J_{i_t})$ (resp., $\rightPoint(J_{i_t})$) may be deployed.
It is easy to verify that except for the good interval $J_{i_t}$ that does not
intersect with $L_t$ and $R_t$, every interval in the stack intersects with
exactly one of these two auxiliary intervals.

The best response of the algorithm would be to output the two auxiliary
intervals and to try to recover the good interval $J_{i_t}$.
(Note that the payoff guaranteed by this strategy is at least $2$ per stack,
whereas any other strategy yields a payoff of at most $2$ per stack.)
For that purpose, the algorithm has to recover the exact locations of the
endpoints of $J_{i_t}$ that implicitly encode $\pi_t(i_t)$.
Observing that the endpoints in this construction can be represented by bit
strings of length $\log(n) + \log(k)$, Theorem~\ref{theorem:LowerBoundProper}
follows by Corollary~\ref{corollary:Gadget}.

%%%%%%%%%%%%%%%%%%%%%%%%%%%%%%%%%%%%%%%
\paragraph{A $2$-lower bound for general intervals.}
%%%%%%%%%%%%%%%%%%%%%%%%%%%%%%%%%%%%%%%
The interval stream that realizes the $(k, n)$-gadget for the $2$-lower
bound for general intervals is constructed as follows.
Assume that $k = 2^{\kappa} - 1$ for some positive integer $\kappa$ and
consider a perfect binary tree $T$ of depth $\kappa$.
The $k$ stacks are identified with the internal nodes of $T$ so that stack $t$
precedes stack $t + 1$ in a pre-order traversal of $T$.
(In other words, if stack $t$ is identified with node $u$ and stack $t'$ is
identified with a child of $u$, then $t < t'$.)
In addition to the intervals in the stacks, we also introduce $2^{\kappa} = k
+ 1$ auxiliary intervals which are identified with the leaves of $T$;
these auxiliary intervals arrive last in the stream.
We say that an interval $J$ is \emph{assigned} to node $u \in T$ if $J$
belongs to the stack identified with $u$ or if $u$ is a leaf and $J$ is the
auxiliary interval identified with it.

The deployment of the stacks and the auxiliary intervals in $\Reals$ is
performed as follows.
Stack $1$ (identified with $T$'s root) is deployed in $[0, 1)$.
Given the deployment of stack $t$ identified with internal node $u \in T$ in
the segment $[x, y)$, we deploy the stacks identified with the left and right
children of $u$ in the segments
$$
\sigma_{\ell}
~ = ~
\left[ x + \lambda (i_t - 3 / 2), x + \lambda (i_t - 1) \right)
\qquad \text{and} \qquad
\sigma_{r}
~ = ~
\left[ x + \lambda (i_t + n - 1 / 2), x + \lambda (i_t + n) \right) ~ ,
$$
respectively, where recall that $\lambda = \frac{y - x}{2 n - 1 / 2}$.
If the children of $u$ are leaves in $T$, then we deploy auxiliary
intervals in those two segments instead of stacks, that is, one auxiliary
interval in $\sigma_{\ell}$ and one in $\sigma_{r}$.
Refer to Figure~\ref{figure:Stack} for illustration.

The key observation regarding the choice of $\sigma_{\ell}$ and $\sigma_{r}$
is that
\begin{align*}
& \leftPoint(J_{i_t - 1}) \leq \leftPoint(\sigma_{\ell}) <
\rightPoint(\sigma_{\ell}) \leq \leftPoint(J_{i_t})
\quad \text{and} \\
& \rightPoint(J_{i_t}) \leq \leftPoint(\sigma_{r}) < \rightPoint(\sigma_{r})
\leq \rightPoint(J_{i_t + 1}) \, .
\end{align*}
In particular, this implies that:
(1) the good interval in the stack identified with node $u \in T$ does not
intersect with any interval assigned to a descendant of $u$ in $T$; and
(2) a non-good interval in the stack identified with node $u \in T$ contains
every interval assigned to a descendant of either the left child of $u$ or the
right child of $u$ in $T$.

Since the $k + 1$ auxiliary intervals are non-intersecting and arrive last in
the stream, they can be included in the output of the algorithm without
requiring any additional space (on top of that dedicated to their
representation).
Moreover, an auxiliary interval intersects with at most one interval in any
valid solution, hence it is a dominant strategy on behalf of the algorithm to
output all the auxiliary intervals.
Therefore, the best response of the algorithm can include an interval $J_i$ of
stack $t$, $1 \leq t \leq k$, in the output only if it is the good interval of
that stack, namely, $i = i_t$.

By definition, in order to include interval $J_{i_{t}}$ of stack $t$ in the
output, the algorithm must hold the exact locations of its endpoints.
The construction of stack $t$ based on permutation $\pi_{t}$ implies that the
exact locations of interval $J_{i_{t}}$'s endpoints encode the value of
$\pi_t(i_t)$.
Observing that the endpoints in this construction can be represented by bit
strings of length $\log(n) \cdot \log(k)$,
Theorem~\ref{theorem:LowerBoundGeneral} follows by
Corollary~\ref{corollary:Gadget}.

%%%%%%%%%%%%%%%%%%%%%%%%%%%%%%%%%%%%%%%%%%%%%%%%%%%%%%%%%%%%%%%%%%%%%%%%%%%%%%
\section{Multiple-Pass Algorithms}
\label{section:MultiPass}
%%%%%%%%%%%%%%%%%%%%%%%%%%%%%%%%%%%%%%%%%%%%%%%%%%%%%%%%%%%%%%%%%%%%%%%%%%%%%%
We extend now the streaming algorithms to use multiple passes through
the data.
First, some notation.
For an interval $I$, let $next(I)$ be the interval
in the input that ends earliest among those that start after $I$
ends, and let $prev(I)$ be the interval that starts latest among
those that finish before $I$ starts. 
We use the notation $next^i(I)$ defined recursively as $I$ when
$i=0$ and as $next(next^{i-1}(I))$ for $i > 0$, and define $prev^i(I)$ similarly.
Observe that if $I$ is available before a pass, 
then a streaming algorithm can easily compute $next(I)$ and $prev(I)$
by the end of the pass, while maintaining $O (1)$ intervals in the memory at
all times.

The multi-pass algorithm runs as follows.  The first pass consists of
the earlier one-pass algorithm, either as the algorithm of
\Section{}~\ref{section:MainAlgorithm} for general intervals, or the 
algorithm of \Section{}~\ref{section:ProperIntervals} 
for proper intervals.
The result of this pass is the set $A$, whichever base algorithm is used.
Let $N_0 = P_0 = A$.  In round $p > 1$, the algorithm
inductively computes $N_{p-1} = \{next(I) : I \in N_{p-2}\}$ and
$P_{p-1} = \{prev(I) : I \in P_{p-2}\}$. Let $A_p = \cup_{i\ge 0}^{p-1} (N_i \cup P_i) = \{next^i(I), prev^i(I) : I
\in A, 0 \le i \le p-1\}$ denote the combined set of 
intervals stored after pass $p$.
When requested, the algorithm produces as output the maximum interval selection
in $A_p$.
This completes the specification of the algorithm.

We first observe that $|A_p| \leq (2 p - 1) A$, hence the space used in phase
$p$ is at most $2 p -1$  times larger than the length of the bit string representing
$A$.

%Define the \emph{span} of a set $R$ of intervals to be the segment given by the
%leftmost and rightmost points in intervals in $R$.

Define the \emph{span} of a set $R$ of intervals to be the segment  with endpoints being 
the leftmost  left endpoint of the intervals in $R$, and the rightmost right endpoint 
of the intervals in $R$.

\begin{lemma} \label{lem:gen-contain}
Given an input of general intervals, the set $A$ computed by the
algorithm \Alg{} of \Section{}~\ref{section:MainAlgorithm} satisfies the
following property:
for any pair of disjoint intervals $I_1$ and $I_2$ in the input,
$A$ contains an interval within the span of $\{I_1, I_2\}$ (given by
$[\lefty(I_1), \righty(I_2))$, assuming $I_1 < I_2$).
\end{lemma}

The following lemmas apply both to general or proper intervals.
An interval is said to be \emph{end-simplicial}  with respect to a set of intervals $X$, 
if it contains either the 
leftmost right endpoint or the rightmost left endpoint of its
connected component with respect to $X$.

\begin{lemma} \label{lem:simplicial}
The set $A$ contains all the end-simplicial intervals with respect to $S$.
\end{lemma}
\begin{proof}
Regarding general intervals,
recall from Proposition~\ref{proposition:VirtualEndpoints} that virtual
intervals in {\Alg} are formed by the intersection of two intervals in the
input.
Thus, if $I$ is end-simplicial with respect to $S$, it contains no virtual interval, and
certainly no actual intervals. Hence, $I$ is admitted to $A$ and never rejected.
For proper intervals, an end simplicial interval on the left (right)
will always represent $R_k$ ($L_k$) for its finishing (beginning) zone $k$.
Thus, it is contained in $A$.
\end{proof}

\begin{lemma} \label{lem:onions}
%Consider the end of an arbitrary round in pass number $p$.
let $I$ be an interval in $A_p$ and let $s \le p-1$.
Let $R \subseteq S$, s.t. $I \in R$ be a set of $s+1$ disjoint intervals.
Then, $A_p$ contains a set of $s + 1$ intervals within the span of $R$.
\end{lemma}
\begin{proof}
Suppose $R$ contains intervals $I,I_1, I_2, \ldots, I_s$ , s.t. $ I_1 < I_2
\ldots I_j <I < I_{j+1} < \ldots < I_s$.
%(The case with intervals on the left of $I$ is symmetric.)
By definition, the intervals $next^i(I)$, $0 \le i \le s-j$, are disjoint
and contained in $A_{s+1}$ and thus also in $A_p$.
Also, by induction, $next^i(I) \le I_i$, for $i=1, \ldots, s-j$,
and thus they fall within the span of $R$.
A similar claim holds for the intervals $prev^i(I)$, $ 0 \leq i  \leq j $
\end{proof}

\begin{lemma} \label{lem:within-span}
Consider any set $R$ of $m$ disjoint intervals in $S$, where 
$m=2p$ for general intervals and $m=2p+1$ for proper intervals.
Then, $A_p$ contains $m-1$ intervals within the span of $R$.
\end{lemma}
\begin{proof}
Follows from Lemma \ref{lem:onions}, along with Lemma
\ref{lem:gen-contain} (resp. Lemma \ref{lem:triple-span}) 
for general (resp. proper) intervals.
\end{proof}

\begin{theorem} \label{theorem:MultiPass}
The multi-pass algorithm finds  a solution for  the interval selection  problem on general
intervals that is a $1 + \frac{1}{2p-1}$-approximation, at the end of  each pass $p$.
On proper intervals it finds a $1 + \frac{1}{2p}$-approximation.
The space used  by the algorithm is $O(p)$ times the size of the output.
\end{theorem}
\begin{proof}
Define $m=2p$ for general intervals and $m=2p+1$ for proper intervals.
Consider an optimal interval selection with intervals $I_1, \ldots,
I_{|OPT|}$, where $\alpha = |OPT|$.
Let $r = \alpha \bmod m$, and $q = \lfloor \alpha/m \rfloor$.
Also let $t = \lceil r/2\rceil$ and $t' = \lfloor r/2\rfloor$.
For each $R_i = \{I_{t + 1 + i m}, \ldots, I_{t+ (i+1)m}\}$, where
$i=0\ldots, q-1$, it holds by Lemma \ref{lem:within-span} that $A_p$ contains
$m-1$ intervals within the span of $R_i$.
By Lemmas \ref{lem:simplicial} and \ref{lem:onions},
$A_p$ also contains $t$ disjoint intervals
within the span of $[\lefty(S),\righty(I_{t})]$ 
and $t'$ disjoint intervals within the span of $[\lefty(I_{m - t'
+1}),\righty(S)]$.
Hence, $A_p$ contains at least 
$q (m-1) + t + t' = \alpha - q \ge \alpha (m-1)/m$ disjoint intervals.
\end{proof}

%%%%%%%%%%%%%%%%%%%%%%%%%%%%%%%%%%%%%%%%%%%%%%%%%%%%%%%%%%%%%%%%%%%%%%%%%%%%%%
\section{Online Algorithm}
\label{section:Online}
%%%%%%%%%%%%%%%%%%%%%%%%%%%%%%%%%%%%%%%%%%%%%%%%%%%%%%%%%%%%%%%%%%%%%%%%%%%%%%
In this section we briefly show how to use the streaming algorithm presented
in \Section{}~\ref{section:MainAlgorithm} to derive a randomized preemptive
online interval selection algorithm.
Our algorithm is $6$-competitive and on top of maintaining at any time
the set of currently accepted intervals $A^{*}$, its only additional memory is
an interval set of cardinality linear in the size of the current optimum.
We thus answer an open question of Adler and Azar~\cite{AdlerAzar03} about the
space complexity of randomized preemptive online algorithms for our problem.

Recall that our streaming algorithm maintains a set $A$ of intervals.
With respect to that set, our algorithm is a deterministic preemptive online
algorithm, adding an interval to $A$ only when that interval arrives, and
possibly preempting it later.
By Corollary \ref{corollary:Approximation}, the cardinality of the set $A$ is
at least half the cardinality of the optimal solution of the input seen so
far.
Moreover, combining Lemma~\ref{lemma:Structural}(P6) and
Lemma~\ref{lemma:Structural}(P7), we conclude that every interval added to $A$
intersects with at most $2$ previous intervals in $A$.
Therefore, $A$ is \emph{online $3$-colorable}:
upon addition into $A$, each interval can be assigned one of three colors,
such that intersecting intervals always have different colors.

Our preemptive algorithm is now simple.
We initially pick a random color $c$ in $\{1, 2, 3\}$.
We then run the streaming algorithm on each received interval $I$, adding $I$
to $A$, and preempting intervals from $A$ as does the streaming algorithm.
If $I$ is added to $A$ we assign it a valid color from $\{1, 2, 3\}$ in a
first-fit manner.
Our solution $ALG$ consists of every interval $J$ in $A$ whose color is $c$.
Clearly, $E[|ALG|] = |A| / 3 \geq |OPT| / 6$, that is, the algorithm is
$6$-competitive.

%%%%%%%%%%%%%%%%%%%%%%%%%%%%%%%%%%%%%%%%%%%%%%%%%%%%%%%%%%%%%%%%%%%%%%%%%%%%%%
\subsection*{Acknowledgments}
%%%%%%%%%%%%%%%%%%%%%%%%%%%%%%%%%%%%%%%%%%%%%%%%%%%%%%%%%%%%%%%%%%%%%%%%%%%%%%
We thank Jaikumar Radhakrishnan and Oded Regev for helpful discussions.

\clearpage

\bibliographystyle{abbrv}
\bibliography{references}

\begin{thebibliography}{10}

\bibitem{AdlerAzar03}
R.~Adler and Y.~Azar.
\newblock Beating the logarithmic lower bound: Randomized preemptive disjoint
  paths and call control algorithms.
\newblock {\em J. Scheduling}, 6(2):113--129, 2003.

\bibitem{Agarwal2010}
P.~K. Agarwal and R.~Sharathkumar.
\newblock Streaming algorithms for extent problems in high dimensions.
\newblock In {\em SODA}, pages 1481--1489, 2010.

\bibitem{AG09}
K.~Ahn and S.~Guha.
\newblock Graph sparsification in the semi-streaming model.
\newblock In {\em ICALP}, pages 328--338, 2009.

\bibitem{AMS99}
N.~Alon, Y.~Matias, and M.~Szegedy.
\newblock The space complexity of approximating the frequency moments.
\newblock {\em J. Comput. Syst. Sci.}, 58(1):137--147, 1999.

\bibitem{ABFR94}
B.~Awerbuch, Y.~Bartal, A.~Fiat, and A.~Ros\'{e}n.
\newblock Competitive non-preemptive call control.
\newblock In {\em SODA}, pages 312--320, 1994.

\bibitem{BHS10}
U.~T. Bachmann, M.~M. Halld{\'o}rsson, and H.~Shachnai.
\newblock Online scheduling intervals and $t$-intervals.
\newblock In {\em SWAT}, 2010.

\bibitem{CP15}
S.~{Cabello} and P.~{P{\'e}rez-Lantero}.
\newblock {Interval Selection in the Streaming Model}.
\newblock {\em ArXiv e-prints}, Jan. 2015.

\bibitem{CanettiI98}
R.~Canetti and S.~Irani.
\newblock Bounding the power of preemption in randomized scheduling.
\newblock {\em SIAM J. Comput.}, 27(4):993--1015, 1998.

\bibitem{clrs}
T.~H. Cormen, C.~E. Leiserson, R.~L. Rivest, and C.~Stein.
\newblock {\em Introduction to Algorithms}.
\newblock MIT Press and McGraw-Hill, 3rd edition, 2009.

\bibitem{EHR12}
Y.~Emek, M.~Halld\'{o}rsson, and A.~Ros\'{e}n.
\newblock Space-constrained interval selection.
\newblock In {\em ICALP}, pages 302--313, 2012.

\bibitem{EpsteinL10}
L.~Epstein and A.~Levin.
\newblock Improved randomized results for the interval selection problem.
\newblock {\em Theor. Comput. Sci.}, 411(34-36):3129--3135, 2010.

\bibitem{EpsteinLMS10}
L.~Epstein, A.~Levin, J.~Mestre, and D.~Segev.
\newblock Improved approximation guarantees for weighted matching in the
  semi-streaming model.
\newblock In {\em STACS}, pages 347--358, 2010.

\bibitem{Feigenbaum2005}
J.~Feigenbaum, S.~Kannan, A.~McGregor, S.~Suri, and J.~Zhang.
\newblock On graph problems in a semi-streaming model.
\newblock {\em Theor. Comput. Sci.}, 348:207--216, December 2005.

\bibitem{FKMSZ08}
J.~Feigenbaum, S.~Kannan, A.~McGregor, S.~Suri, and J.~Zhang.
\newblock Graph distances in the data-stream model.
\newblock {\em SIAM J. Comput.}, 38(5):1709--1727, 2008.

\bibitem{FPZ08}
S.~Fung, C.~Poon, and F.~Zheng.
\newblock Improved randomized online scheduling of unit length intervals and
  jobs.
\newblock In {\em WAOA}, 2008.

\bibitem{Gavril72}
F.~Gavril.
\newblock Algorithms for minimum coloring, maximum clique, minimum covering by
  cliques, and maximum independent set of a chordal graph.
\newblock {\em SIAM J. Comput.}, 1(2):180--187, 1972.

\bibitem{HalldorssonHLS10}
B.~V. Halld{\'o}rsson, M.~M. Halld{\'o}rsson, E.~Losievskaja, and M.~Szegedy.
\newblock Streaming algorithms for independent sets.
\newblock In {\em ICALP}, pages 641--652, 2010.

\bibitem{HRR98}
M.~R. Henzinger, P.~Raghavan, and S.~Rajagopalan.
\newblock Computing on data streams.
\newblock In {\em AMS-DIMACS series, special issue on computing on very large
  datasets}. 1998.

\bibitem{Kleinberg05}
J.~Kleinberg and E.~Tardos.
\newblock {\em Algorithm Design}.
\newblock Addison Wesley, 2005.

\bibitem{LMPR01}
S.~Leonardi, A.~Marchetti-Spaccamela, A.~Presciutti, and A.~Ros\'{e}n.
\newblock On-line randomized call control revisited.
\newblock {\em SIAM J. Comput.}, 31(1):86--112, 2001.

\bibitem{Lipton94onlineinterval}
R.~J. Lipton and A.~Tomkins.
\newblock Online interval scheduling.
\newblock In {\em SODA}, pages 302--311, 1994.

\bibitem{McGregor05}
A.~McGregor.
\newblock Finding graph matchings in data streams.
\newblock In {\em APPROX-RANDOM}, pages 170--181, 2005.

\bibitem{MunroPaterson80}
J.~I. Munro and M.~Paterson.
\newblock Selection and sorting with limited storage.
\newblock {\em Theor. Comput. Sci.}, 12:315--323, 1980.

\bibitem{Muthukrishnan05}
S.~Muthukrishnan.
\newblock Data streams: Algorithms and applications.
\newblock {\em Foundations and Trends in Theoretical Computer Science}, 1(2),
  2005.

\bibitem{Woeginger94}
G.~J. Woeginger.
\newblock On-line scheduling of jobs with fixed start and end times.
\newblock {\em Theor. Comput. Sci.}, 130(1):5--16, 1994.

\end{thebibliography}

%%%%%%%%%%%%%%%%%%%%%%%%%%%%%%%%%%%%%%%%%%%%%%%%%%%%%%%%%%%%%%%%%%%%%%%%%%%%%%
%%%%%%%%%%%%%%%%%%%%%%%%%%%%%%%%%%%%%%%%%%%%%%%%%%%%%%%%%%%%%%%%%%%%%%%%%%%%%%
\clearpage

\pagenumbering{roman}
\appendix

\renewcommand{\theequation}{A-\arabic{equation}}
\setcounter{equation}{0}

\begin{center}
\textbf{\large{APPENDIX}}
\end{center}

%%%%%%%%%%%%%%%%%%%%%%%%%%%%%%%%%%%%%%%%%%%%%%%%%%%%%%%%%%%%%%%%%%%%%%%%%%%%%%
\section{Lifting the Distinct Endpoints Assumption}
\label{section:LiftingAssumptions}
%%%%%%%%%%%%%%%%%%%%%%%%%%%%%%%%%%%%%%%%%%%%%%%%%%%%%%%%%%%%%%%%%%%%%%%%%%%%%%
Recall that our analysis assumes that all the intervals in the stream $S$ are
closed and that their endpoints are distinct.
In this section we show that these assumptions can be lifted.
A quick glance at our algorithm reveals that it is essentially
\emph{comparison-based}, namely, it can be implemented via a \emph{comparison
oracle} $\Oracle : \Reals^2 \rightarrow \{-1, 0, +1\}$ without accessing the
interval's endpoints in any other way;
given two endpoints $p, q$ of intervals in $S$, the comparison oracle returns
$$
\Oracle(p, q)
~ = ~
\left\{
\begin{array}{ll}
-1 & \text{ if } p < q \\
0 & \text{ if } p = q \\
+1 & \text{ if } p > q ~ .
\end{array}
\right.
$$
The assumption that all endpoints are distinct means that the algorithm and
its analysis rely on a comparison oracle $\Oracle' : \Reals^2 \rightarrow
\{-1, 0, +1\}$ with the additional guarantee that $\Oracle'(p, q) \neq 0$
whenever $p \neq q$.
We shall refer to such a comparison oracle $\Oracle'$ as a
\emph{distinct-endpoints comparison oracle}.

We show that for every stream $S$ of intervals (the endpoints of these
intervals may be arbitrarily open or closed) associated with a comparison
oracle $\Oracle$, there exists a distinct-endpoints comparison oracle
$\Oracle'$ such that for every two intervals $I, J \in S$, the closure of $I$
and the closure of $J$ intersect under $\Oracle'$ if and only if $I$ and $J$
intersect under $\Oracle$.
Moreover, given an access to the comparison oracle $\Oracle$, the
distinct-endpoints comparison oracle $\Oracle'$ can be implemented under our
streaming model's space requirements.

The distinct-endpoints comparison oracle $\Oracle'$ is designed as follows.
Consider an endpoint $p$ of an interval $I \in S$ and an endpoint $q$ of an
interval $J \in S$, $I \neq J$.
If $\Oracle(p, q) \neq 0$, then we set $\Oracle'(p, q) = \Oracle(p, q)$, so
assume hereafter that $\Oracle(p, q) = 0$.
Consider first the case in which $p$ is a right endpoint and $q$ is a left
endpoint (the converse case is analogous).
If at least one of the endpoints is open, then set $\Oracle'(p, q) = -1$;
otherwise (both endpoints are closed), set $\Oracle'(p, q) = +1$.

Now, consider the case in which both $p$ and $q$ are left endpoints (the
converse case is analogous).
If $p$ is open and $q$ is closed, then set $\Oracle'(p, q) = +1$;
if $p$ is closed and $q$ is open, then set $\Oracle'(p, q) = -1$;
if both $p$ and $q$ are open or both are closed, then we set
$$
\Oracle'(p, q)
~ = ~
\left\{
\begin{array}{ll}
+1 & \text{ if $I$ (the interval of $p$) arrived before $J$ (the interval of
$J$)} \\
-1 & \text{ if $I$ (the interval of $p$) arrived after $J$ (the interval of
$J$).}
\end{array}
\right.
$$

It is easy to verify that the closures of every two intervals intersect under
$\Oracle'$ if and only if the intervals themselves intersect under $\Oracle$.
Therefore, it remains to show that $\Oracle'$ can be implemented in the
streaming model.
Apart from an access to the original comparison oracle $\Oracle$, the
implementation of $\Oracle'(p, q)$ is based on:
(1) knowing for each endpoint whether it is a left endpoint or a right
endpoint;
(2) knowing for each endpoint whether it is open or closed; and
(3) knowing the order of arrival of intervals that share a left (respectively,
right) endpoint.
The first two requirements are clearly satisfied by the information provided
in the input.
For the third requirement, we note that if two intervals share a left (resp.,
right) endpoint $p$, then they must intersect.
Thus, Lemma~\ref{lemma:Structural}(P5) and Lemma~\ref{lemma:Structural}(P6)
guarantee that at any given time, our algorithm maintains $O (1)$ intervals
that have $p$ as their left (resp., right) endpoint.
A data structure that tracks the arrival order of these intervals can
therefore be implemented with $O (1)$ additional bits per interval.

%%%%%%%%%%%%%%%%%%%%%%%%%%%%%%%%%%%%%%%%%%%%%%%%%%%%%%%%%%%%%%%%%%%%%%%%%%%%%%
\section{Proof of Lemma~\ref{lemma:ImpossibleCode}}
\label{section:ProofOfLemmaImpossibleCode}
%%%%%%%%%%%%%%%%%%%%%%%%%%%%%%%%%%%%%%%%%%%%%%%%%%%%%%%%%%%%%%%%%%%%%%%%%%%%%%
Let $n$ be sufficiently large so that $n (1 + \log(e)) \leq \alpha
n \log(n)$.
Suppose toward a contradiction that there exist two functions $\hat{f} : P_n
\rightarrow \{0, 1\}^{m}$ and $\hat{g} : \{0, 1\}^{m} \times [n] \rightarrow
[n]$ such that $\Probability( \hat{g}(\hat{f}(\pi), i) = \pi(i)) \geq 2
\alpha$.
We shall use these functions to construct a uniquely decodable coding scheme
$s : P_n \rightarrow \{0, 1\}^{*}$ so that $\Expectation_{\pi \in_{r} P_n}
[|s(\pi)|] < \log(n!)$.
This contradicts Shannon's source coding theorem as the entropy of choosing
$\pi$ uniformly at random from $P_n$ is $\log(n!)$.

In order to construct the coding scheme, we first define the vector $v_{\pi}
\in \{0, 1\}^{n}$ for every $\pi \in P_n$ by setting $v_{\pi}(i) = 1$ if
$\hat{g}(\hat{f}(\pi), i) = \pi(i)$; and $v_{\pi}(i) = 0$ otherwise.
Let $W_{\pi} = \{ i \in [n] \mid v_{\pi}(i) = 0 \}$.
The coding scheme $s$ is now defined by setting the codeword of each $\pi \in
P_n$ to be
$$
s(\pi)
~ = ~
v_{\pi} \circ \hat{f}(\pi) \bigcirc_{i \in W_{\pi}} \pi(i) ~ ,
$$
where $\bigcirc_{i \in W_{\pi}} \pi(i)$ denotes a concatenation of the
standard binary representations of $\pi(i)$ for all $i \in W_{\pi}$ listed in
increasing order of the index $i$.

We first argue that $s$ is indeed a uniquely decodable code.
To that end, notice that for every $\pi \in P_n$ and for every $i \in [n]$, we
can extract the value of $\pi(i)$ from $s(\pi)$ as follows: \\
(1) Check in $v_{\pi}$ if the correct value of $\pi(i)$ can be extracted from
$\hat{f}(\pi)$, that is, if $v_{\pi}(i) = 1$. \\
(2) If it can ($v_{\pi}(i) = 1$), then $\pi(i)$ is extracted by computing
$\hat{g}(\hat{f}(\pi), i)$ (recall that $\hat{f}(\pi)$ is found in the second
segment of $s(\pi)$). \\
(3) Otherwise ($v_{\pi}(i) = 0$), $\pi(i)$ is extracted from the third segment
of $s(\pi)$. \\
Moreover, the coding scheme $s$ is prefix-free (and hence uniquely decodable)
since $v_{\pi} = v_{\pi'}$ implies that $|s(\pi)| = |s(\pi')|$ for every two
permutations $\pi, \pi' \in P_n$.
Thus, if the codewords $s(\pi)$ and $s(\pi')$ agree on the first $n$ bits,
then they must have the same length, which means that $s(\pi)$ cannot be a
proper prefix of $s(\pi')$.

It remains to show that $\Expectation_{\pi \in_{r} P_n} [|s(\pi)|] < \log(n!)$.
By definition, $|s(\pi)| = n + m + \log(n) \cdot |W_{\pi}|$ for every $\pi
\in P_n$, so
$$
\Expectation_{\pi \in_{r} P_n} [|s(\pi)|]
~ = ~
n + m + \log(n) \cdot \Expectation_{\pi \in_{r} P_n} [|W_{\pi}|] ~ .
$$
The assumption that $\Probability_{\pi \in_{r} P_n, i \in_{r}
[n]}(\hat{g}(\hat{f}(\pi), i) = \pi(i)) \geq 2 \alpha$ implies that
$\Probability_{\pi \in_{r} P_n, i \in_{r} [n]}(i \in W_{\pi}) \leq 1 - 2
\alpha$, hence $\Expectation_{\pi \in_{r} P_n} [|W_{\pi}|] \leq (1 - 2 \alpha)
n$.
Plugging $m = \alpha n \log(n)$, we conclude that
$$
\Expectation_{\pi \in_{r} P_n} [|s(\pi)|]
~ \leq ~
n + (1 - \alpha) n \log(n) ~ .
$$
By the choice of $n$ (satisfying $n (1 + \log(e)) \leq \alpha n \log(n)$), we
derive the desired inequality since $\log(n!) > n \log(n) - n \log(e)$.
The assertion follows.

\end{document}